\pdfoutput=1
\documentclass[11pt]{JHEP3}
\input epsf.tex
\epsfclipon

\usepackage{color}
\let\normalcolor\relax
\usepackage{epsfig}
\usepackage{epstopdf}
\usepackage{epsf}
\input{epsf.sty}
\usepackage{graphicx,amsmath,amssymb}
\usepackage{epsfig,multicol}
\usepackage{multirow}
\allowdisplaybreaks

\usepackage{bbm,bm,amsmath,amssymb}
 
\usepackage[normalem]{ulem}

\usepackage{comment}
\def\be{\begin{equation}}
\def\ee{\end{equation}}
\def\figs/B{B}
\def\bea{\begin{eqnarray}}
\def\eea{\end{eqnarray}}
\def\bg{\begin{eqnarray}}
\def\nd{\end{eqnarray}}

\def\log{{\rm log}}

\title{Quantum Corrections and the de Sitter Swampland Conjecture}
\author{Keshav Dasgupta$^{1}$, Maxim Emelin$^{1}$, Evan McDonough$^{2}$,  
	Radu Tatar$^{3}$\\
	\vskip.03in
	${}^1$ Department of Physics, McGill University, Montr\'{e}al, Qu\'{e}bec, H3A 2T8, Canada \\
	${}^2$ Department of Physics, Brown University,  Providence, RI, 02906, USA \\
	${}^3$ Department of Mathematical Sciences,
University of Liverpool,  Liverpool, L69 7ZL, United Kingdom \\
	{\tt keshav@hep.physics.mcgill.ca, maxim.emelin@mail.mcgill.ca}
	~~{\tt  evan${}_-$mcdonough@brown.edu, Radu.Tatar@Liverpool.ac.uk}}

\date{\today}
\maketitle

\abstract{Recently a swampland criterion has been proposed that rules out de Sitter vacua in string theory. Such a criterion should hold at all points in the field space and especially at points where the system is on-shell. However there has not been any attempt  to examine the swampland criterion against explicit equations of motion. In this paper we study four-dimensional de Sitter and quasi-de Sitter solutions using dimensionally reduced M-theory. While on one hand all classical sources that could allow for solutions with de Sitter isometries are ruled out, the quantum corrections, on the other hand, are found to allow for de Sitter solutions provided certain constraints are satisfied. A careful study however shows that generically such a constrained system does not allow for an effective field theory description in four-dimensions. Nevertheless, if some hierarchies between the various quantum pieces could be found, certain solutions with an effective field theory description might exist. Such hierarchies appear once some mild
 time dependence is switched on, in which case certain quasi-de Sitter solutions may be found without a violation of the swampland criterion.}

\begin{document}

\section{Introduction}	
\label{sec:intro}

There is a long history of no-go theorems for de Sitter solutions in string theory. Starting with the supergravity work of Gibbons \cite{gibbons} and continued by Maldacena-Nunez \cite{malnun}, no-go theorems for dS have been formulated with ever increasing breadth \cite{nogo, Kutasov:2015eba, Green:2011cn, hirano, Sethi:2017phn, Danielsson} (see also \cite{wittendS, gates} for related discussions). These works led to the recent proposal \cite{vafa1} that four-dimensional theories derived from string theory, i.e.~the string landscape, satisfy a universal bound, referred to as the \emph{de Sitter swampland conjecture}:
\be
{\vert\nabla V\vert \over V} \geq c  .
\label{dSSC}
\ee
The conjecture states that any four-dimensional effective field theory which violates this bound does not have an embedding in string theory, and hence is in the so-called swampland \cite{Brennan:2017rbf,Ooguri:2006in}. Since \cite{vafa1}, a number of works have studied and extended the above conjecture \cite{swampland,Achucarro:2018vey,Heisenberg:2018yae}.

The related \emph{swampland distance conjecture} \cite{Ooguri:2006in} states that the range traversed by scalar fields in field space is bounded by $\Delta \sim \mathcal{O}(1)$ in Planck units. More quantitatively, the conjecture asserts that at large field excursion $D$ that there emerges a tower of light states with mass given by:
\be
\label{dSSC2}
m \sim M_{p} e^{- \alpha D},
\ee
where $\alpha$ is an order 1 number. This implies the breakdown of low energy effective field theory, as  can be seen for simple examples such as a Kaluza-Klein reduction on a circle, in the limit when the circle becomes large.

The combination of these two conjectures leads to interesting possibilities. As discussed in  \cite{vafa2}, it is easy to construct cosmological toy models that satisfy the swampland conjecture \eqref{dSSC} but violate the distance conjecture. It is also possible to construct cosmological toy models that satisfy both criteria. The prime examples of these cases are large field inflation and quintessence respectively.

However, it remains to be determined what combination of these conjectures is realized by string theory. In this context, two questions arise regarding the fate of the dS in string theory:
 \begin{enumerate}
\item Do there exist explicit solutions to the ten dimensional equations of motion that violate \eqref{dSSC}? If so, do they have descriptions as four dimensional effective field theories?
\item Do there exist complicated field configurations which realize an exact or quasi $dS_4$ \emph{without} violating \eqref{dSSC}? And again, if so, do they have a corresponding four dimensional effective field theory?
\end{enumerate}
In this work we study these questions by considering the explicit solutions to the ten-dimensional equations of motion, working primarily in the 11-dimensional M-theory lift of $dS_4 \times X_6$ in type IIB string theory, with $X_6$ an arbitrary six-dimensional manifold.

As a prelude to this, we first consider generalities of the de Sitter swampland conjecture. It can be straightforwardly extended to multifield models, solitonic solutions, and time-dependent field configurations, and in all but the latter dS solutions are clearly ruled out by the conjecture. Motivated by this, we consider multi-field models with time-dependence, which under certain conditions can be mapped to higher derivative single-field models. These models allow for positive cosmological constant solutions  \emph{without} violating the swampland conjecture \eqref{dSSC}, but there are no known embeddings of these models in string theory. Despite that, these class of examples suggest the existence of a broader picture in which time-dependent backgrounds in the string landscape 
would form the cornerstone to study cosmological evolution of our universe. 

We then consider the leading $\alpha'$ corrections to type IIB string theory. These enter the ten-dimensional action as higher derivative terms, and manifest themselves in four-dimensions as corrections to the Kahler potential of the multifield model. At leading order in $\alpha'$, no static solutions exist, and hence solutions are intrinsically time-dependent.  We demonstrate that the swampland conjecture \eqref{dSSC} is \emph{never} violated in this setup, and that any solution will eventually decompactify, implying that again the solution ceases to be in the regime of four-dimensional effective field theory.

With these preliminary investigations in mind, we then undertake a more thorough analysis of the effect of string corrections to supergravity. Our main tool in this analysis will be a parametrization of the higher derivative corrections to the supergravity action which arise from the $\alpha'$ and string loop corrections. When lifted to M-theory, this manifests itself as a complicated system of time-dependent equations of motion, of which one can make a surprising amount of sense. This analysis confirms the intuition built by the previous sections that the main players in assessing the dS swampland conjecture in string theory are: (i) higher derivative corrections, (ii) multiple fields, and (iii)  time-dependences\footnote{It is instructive to speculate on the contribution of classical sources, especially O-planes, in generating de Sitter vacua in string theory. The presence of O-planes essentially involve two regions, one, away from the locations of the O-planes, and two, at the locations of the O-planes. The latter involve singular points and one can remove the singular points from the manifold leaving holes, but then certain boundary terms defined in \cite{nogo} become non-zero. To obtain the exact values of these terms, one needs to know the metric near the singularity, but since classical gravity breaks down, we cannot evaluate them without incorporating quantum corrections.}.

The analysis from M-theory provides evidence, though not definitive conclusions, for the answers to the above questions: we find that under certain conditions, there do exist $dS_4$ solutions that violate conjecture \eqref{dSSC}, but they involve a tower of quantum corrections and hence are not in the regime of four-dimensional low-energy effective field theory. If one allows for a time-dependence of the cosmological constant, a so-called \emph{quasi}-de Sitter space, then it is possible to both satisfy the conjecture \eqref{dSSC} and be a valid effective field theory, consistent with the claims regarding quintessence in \cite{vafa2}. While more work is certainly needed to solidify these results, we take this as evidence that fate of dS in string theory is not yet sealed.

The outline of this paper is as follows: In section II we consider implications and generalizations of the swampland conjecture \eqref{dSSC}, and section III we consider toy models with time-dependence. In Section IV we consider the leading $\alpha'$ correction to type IIB string theory, and in section V undertake a more complete analysis via a lift to M-theory. We conclude in Section VI with directions for future work.

\section{Bounds on the potential and the swampland conjecture \label{swampland}}

In string and M-theory the potentials to the moduli fields appear from fluxes and non-perturbative terms in the action. For simplicity let us first assume that there is one moduli field $-$ we will call it $\varphi(x)$ $-$ whose potential may be written as $V(\varphi)$ and $x$ denotes a generic point in the four-dimensional spacetime. 
This potential will appear in the dimensionally reduced action and typically one studies the region near the minima of the potential to discuss fluctuations of the moduli scalar. For the present case, let us assume that we are {\it not} in the minima of the potential such that $\phi$ will denote any generic point in the potential. Let $b(x)$ denotes the local neighborhood of the potential such that:
\bg\label{chukka}
\varphi(x) \equiv \phi(x) + b(x), \nd
where we impose no constraint on $b(x)$ at this stage. Using \eqref{chukka} one can easily show that there exists the following upper bound:
\bg\label{water}
{\vert V(\phi \pm b) - V(\phi)\vert \over V(\phi \pm b)} ~\le ~ 1, \nd  
where we have assumed\footnote{We will be using the symbol $\vert .... \vert$ to denote both the modulus of a number or a function, and the magnitude of a vector, unless mentioned otherwise. Which is which should be clear from the context.} 
that $V(\varphi) > 0$ because we want to study non-supersymmetric vacua.  The $\pm$ sign denotes the choice between locally monotonically increasing or decreasing potentials\footnote{As mentioned, we only demand this 
locally. Globally all we want is $V(\phi) > 0$. There could be inflection point but the analysis will be away from local minima. For example we can  choose $\phi$ at the inflection point and $b$ on either side in such a way that we can demand either 
$V(\phi + b) > V(\phi)$ or $V(\phi - b) > V(\phi)$.}. For example, with locally monotonically 
decreasing potential, 
we can express \eqref{water} with the minus sign where  $V(\phi -b) > V(\phi)$ and 
$V(\phi)/V(\varphi)$ denotes deviation from identity. The inequality \eqref{water} now leads to the following natural bound:
\bg\label{drishyam}
{\left\vert \partial_\phi V(\phi) \mp {b \over 2}~ \partial^2_{\phi ~} V(\phi) \pm {b^2\over 3!}~ \partial^3_{\phi~} V(\phi) + ....\right\vert \over V(\phi \pm b)} ~\le ~ {1\over \vert b \vert}. \nd
Now consider $b(x) \ll 1$, for all points in the four-dimensional spacetime parametrized by $x$. In units used here to write $\phi$ and $b$, this means we take $\phi(x) \gg b(x)$ at all points in the four-dimensional space parametrized by $x$. In this limit, \eqref{drishyam} takes the following upper bound:
\bg\label{te3n}
{\vert \partial_\phi V \vert \over V} ~ \le ~ {1\over \vert b \vert}, \nd
with the assumption that $V \equiv V(\phi)$. Clearly when $b(x) < 1$, but $\phi(x) \gg b(x)$, we can modify the bound 
\eqref{te3n} by keeping the next order in the derivative expansion in the following way:
\bg\label{mesmer}
{\left\vert \partial_\phi V \mp {b\over 2} ~ \partial^2_{\phi~} V\right\vert \over V} ~ \le ~ {1\over \vert b \vert}, \nd
which differs from \eqref{water} by the choice $V(\phi)$ in the denominator as against $V(\phi + b)$. Generically it is clear that, in such a scenario, the upper bound is always:
\bg\label{polley}
{\left\vert \left({{\rm exp}\left(\pm b\partial_\phi\right) - 1}\right)V(\phi) \right\vert \over \vert V(\phi)\vert} ~ \le ~ {1}, \nd
irrespective of the sign of $V(\phi)$. The next question however is whether there exists a \emph{lower} bound to the above ratio. Interestingly, for the case where $b(x) \ll 1$ i.e when the ratio satisfies \eqref{te3n}, the sign ambiguity doesn't appear and precisely for this case a recent work \cite{vafa1} suggested the existence of a lower bound, i.e. the de Sitter swampland conjecture. 

Putting things together, there seems to be the following range for the ratio 
$\vert\partial V\vert/V$:
\bg\label{hurley}
c ~\le ~ {\vert \partial_\phi V\vert \over V} ~ \le {1\over \vert b \vert}, \nd
where $c$ is defined for various cases in \cite{vafa1}. This also means that $b$ cannot be arbitrarily large. Of course increasing $b$ beyond the small neighborhood already means going beyond the first derivative in 
\eqref{hurley}. This then raises the question of the validity of the lower bound beyond the first derivative, although there has been some speculations that the lower bound remains valid even if we include the second derivative \cite{vafa22}. However, in the way we have presented our analysis, there is a sign ambiguity accompanying the  second derivative piece. To avoid such ambiguities we will take $b$ arbitrarily small so that the bound will simply be \eqref{hurley}. \newline 


\noindent {\bf Multifield Generalization: }   For the case with multiple moduli, the field gradient of the potential may be generalized to be a norm  using the metric $g_{\alpha\beta}$ on the moduli space. To see how this works out, let us ask how the lower bound looks like for the case with multiple fields. For concreteness we will label the small deviation of the field $\phi_i$ 
by $b_i$. The condition that we are now looking for is:
\bg\label{mcawin}
c_o ~ \le ~ {\left\vert \left(b\cdot \partial_\phi\right) V\right\vert \over V}, \nd 
 where $c_o$ is the lower bound and we have defined $V$, that appears in both the numerator and the denominator in \eqref{mcawin}, as $V\left(\{\phi_i\}\right)$. Additionally, the way 
  we have defined    
  $\left(b\cdot \partial_\phi\right) V$, one may easily impose the following inequality: 
 \bg\label{bavalace} 
  \vert\left(b\cdot \partial_\phi\right) V\vert \equiv \Big\vert\sum_\alpha b_\alpha \partial_{\phi^\alpha} V\Big\vert 
  ~ < ~ \vert b\vert \vert \nabla V\vert, \nd 
  where we have used Cauchy-Schwartz theorem to show that the the product of the components of two vectors is 
  smaller than the product of the magnitudes of the two vectors\footnote{We can take the two vectors as
  $b \equiv \sum_\alpha b_\alpha \hat{\varphi}_\alpha$ and $\nabla V \equiv \sum_\beta \partial_{\phi^\beta} V\hat{\varphi}_\beta$ with $\hat{\varphi}_\alpha $ being the unit vectors in the field space that are not necessarily orthogonal. The magnitudes may then be defined as $\vert b\vert \equiv \sqrt{g^{\alpha\beta}b_\alpha b_\beta}$  and
  $\vert\nabla V\vert \equiv  \sqrt{g^{\alpha\beta} \partial_\alpha V \partial_\beta V}$ using the metric $g_{\alpha\beta}$ 
  in the moduli space. \label{kochi}}.
  This immediately tells us that the lower bound in the presence of multiple fields
  can be expressed as:
  \bg\label{kinsella}
  {c_o\over \vert b\vert} ~ \le ~ {\vert\nabla V\vert \over V}, \nd
  where $\vert b\vert$ and $\vert \nabla V\vert$ are defined in the way described earlier, 
  and one may express the left hand side of the inequality 
  \eqref{kinsella} simply as $c$ of \cite{vafa1}. \newline 
    
\noindent {\bf Solitons:} The way we have defined $\partial_\phi V$ includes not only the constant value of 
$\phi$ but also solitonic solutions. This implies that the moduli of the internal manifold, in either string theory or M-theory, could be functions of the three spatial coordinates of the four-dimensional spacetime parametrized by $x$ above. For example we could take $V$ as:
\bg\label{icmp}
V(\phi) \equiv {1\over 2} \left(\partial_{\bf r} \phi\right)^2 + U(\phi), \nd
where $U(\phi)$ is typically a polynomial,  with or without higher derivatives, in $\phi$, and we have defined 
$x \equiv ({\bf r}, t)$. 
This way all kinds of {\it static} solutions are covered, implying that for a vacua with positive cosmological constant to be a solution we will require:
\bg\label{simenon}
\partial_\phi V \equiv {\delta V\over \delta \phi} =  - \left(\partial_{\bf r} \cdot \partial_{\bf r}\right) \phi + \partial_\phi U = 0, ~~~~ V(\phi) > 0, \nd
violating the lower bound\footnote{In other words, at the point where $\phi$ satisfies equation of motion, $c = 0$, thus violating the bound. The swampland conjecture must hold at {\it all} points in the field space irrespective of whether $\phi$ satisfies equation of motion or not. For example let us consider the simple case with 
$V = {1\over 2} m^2\phi^2$. At any generic point in the landscape, it is easy to see that 
${\vert \partial_\phi V\vert \over V} = {1\over \vert\phi\vert}$. When 
$\phi$ satisfies EOM with static solution i.e with $\phi = 0$, the swampland bound is trivially satisfied. However issue arises when $\phi$ becomes arbitrarily large in the landscape. This is avoided by the appearance of new degrees of freedom, changing the very potential that we started off with (see \cite{vafa1} for more details).} of the range given in \eqref{hurley}. Thus if \eqref{hurley} is the allowed range for the ratio
$\vert\partial V\vert/V$, de Sitter vacua are completely excluded from the system. \newline 

\noindent {\bf Time-Dependent Fields:} A caveat to the above reasoning is if time dependence is involved where, assuming $\phi$ to be a canonical scalar field,
\bg\label{vidya}
{\vert\partial_\phi V\vert\over V} ~ =  ~ {\vert 3H \dot\phi + \ddot{\phi} \vert \over V(\phi)} ~ \ge ~ c,  \nd
with $H$ and the dots denoting the Hubble constant and the time derivatives respectively. 
Generically such time dependences of the moduli fields lead to time dependent Newton's constant in four-dimensional spacetime and is therefore disfavored. However if the time variations are slow on cosmological scales, this may not be an issue. On the other hand a time variation like \eqref{vidya} does not necessarily imply a de Sitter vacua so we will have to tread carefully. For example with a single scalar field, and assuming the lower bound in \eqref{vidya} being saturated, we get a potential:
\bg\label{katey} 
V(\phi) = \vert A\vert e^{c\phi}, \nd
where $A$ is a constant and $c$ is the lower bound in \eqref{vidya}. For time-dependent $\phi$, this implies a time-dependent potential which in turn is related to 
a time varying cosmological constant. On the other hand, if we have multiple fields, the fields $\phi_i$ can be time-dependent but the potential $V\left(\{\phi_k\}, t\right)$, which can now be viewed along with a quantum piece, can be made time-independent.  In general however we expect the following inequality (for $\partial_0 V > 0$):
\bg\label{jlilharmony}
\Big\vert \sum_i \dot{\phi}_i \partial_{\phi_i} V\left(\{\phi_k\}, t \right)\Big\vert  + \left\vert{\partial V\left(\{\phi_k\}, t\right) \over \partial t}\right\vert ~\ge~  \left\vert {dV\left(\{\phi_k\}, t \right)\over dt}\right\vert ~\ge ~  0,  \nd
where an equality in the last sequence will lead to the required time-independent case. Note that the above inequality is generic and 
using similar arguments as in \eqref{bavalace}, we can rewrite \eqref{jlilharmony} as the following series of inequalities:  
\bg\label{jenga}
\Big\vert \sum_i {\dot\phi}_i \hat{\varphi}_i\Big\vert \Big\vert \sum_k \partial_{\phi_k}V \hat{\varphi}_k\Big\vert  ~ > ~ 
\Big\vert \sum_i \dot{\phi}_i \partial_{\phi_i} V\left(\{\phi_k\}, t \right)\Big\vert ~\ge  
- \left\vert{\partial V\left(\{\phi_k\}, t\right) \over \partial t}\right\vert  \nd
where we have again used the Cauchy-Schwartz theorem for the two vectors $\sum_i {\dot\phi}_i \hat{\varphi}_i$ and  
$\sum_k \partial_{\phi_k}V \hat{\varphi}_k$ as we had in \eqref{bavalace} (see also footnore \ref{kochi}). 
The right hand side of the inequality is defined with respect to $\partial V / \partial t$ and so provides a lower bound. We can further rearrange \eqref{jenga} to take the following suggestive form:
\bg\label{cmtota}
{\vert \nabla V\vert \over V} ~ \ge ~ - {1\over V~ \sqrt{g^{ab} \partial_0{\phi}_a \partial_0{\phi}_b}} 
\left\vert {\partial V\over \partial t} \right\vert, \nd
which surprisingly looks like the bound proposed in \cite{vafa1} if we can identify the right hand side of the inequality to $c$ of \cite{vafa1}. There is however no compelling reason for doing so, and therefore we will view \eqref{cmtota} simply as an algebraic identity at this stage\footnote{For example if $\phi_\alpha$ are almost static and the potential 
$V$ has no inherent time dependence then the right hand side of the inequality \eqref{cmtota} vanishes allowing solutions with de Sitter isometries to exist. However the bound in \cite{vafa1} will still forbid such solutions  in the landscape.}. We have also used $g_{ab}$ to denote the metric in the moduli space, exactly as in \eqref{bavalace} before, and should not be confused with the spacetime metric $g_{ij}$ to be used later. It is interesting to note that, if there is no explicit time dependence of $V$, the right hand side will vanish and the inequality will take a simpler form. 
Whether this can always hold is a matter of concern of course, but the very fact that we seem to be getting an inequality close to the one proposed in \cite{vafa1} is intriguing enough.  Our simple exercises presented here however demonstrated that the swampland criteria presented in \cite{vafa1, Ooguri:2006in, vafa22}
are not simple consequences of algebraic identities of the fields. Instead they are based on deeper fundamental principles. This will become clearer when we will study an on-shell example from M-theory in section \ref{mtheory}.
\newline 

\noindent {\bf Non-Supersymmetric Solutions:} Finally, we note that while the lower bound in \eqref{hurley} rules out non-supersymmetric  vacua with positive cosmological constants, it does not rule out vacua that are non-supersymmetric but have {\it zero} cosmological constants. Such vacua arise from compactifications with non-supersymmetric (1, 2) Imaginary Self-Dual (ISD) fluxes \cite{GKP, anke1}. 

In the following we will start with two warm-up examples related to backgrounds with positive cosmological constants before delving ourselves into a detailed study of the consequences from equations of motion in a dimensionally reduced M-theory set-up.   

\section{Towards a background with positive cosmological constant in the landscape}

\label{sec:PX}

The previous section argued that the time-dependent fields may allow for 
positive cosmological constant solutions\footnote{By positive cosmological constant solutions, we will always means solutions for $V > 0$ {\it without} all the de Sitter isometries. For example in section \ref{mtheory} we will see an example where the four-dimensional metric is de Sitter like but the fluxes break the de Sitter isometries.} 
without a violation of the swampland conjecture. Similarly, recent work \cite{Achucarro:2018vey} has argued that multi-field models have the potential for allowing a quasi-dS solution, i.e. inflation, without a violation of the swampland conjecture.  Here we demonstrate the relation between these approaches, and argue that vacua with positive cosmological constants are indeed possible without a violation of the swampland conjecture, and without invoking the equality in the last sequence of  \eqref{jlilharmony}. This provides a way of using \eqref{jenga} for realizing positive cosmological constant solutions from time-dependent fields.  Interestingly, doing so leads to higher derivative terms in the single field effective action.

We start with the action of a non-linear $\sigma$ model,
\be
\mathcal{L} = \sqrt{-g} \left[  - G^{I J}( \phi ) \partial_\mu \phi^I \partial^\mu \phi^J - V(\phi ) \right]
\ee
where $\phi^I$ are a set of $N$ scalars, $G^{IJ}$ is the metric on the kinetic manifold, and $V(\phi)$ is the potential, both of which in principle depend on all the $\phi^I$. For simplicity we work with a two-field example where only one of the fields has a non-canonical kinetic term, and the potential depends only on this field. We consider:
\be
\mathcal{L} = \sqrt{-g} \left[  - \frac{1}{2} f(\chi) (\partial \phi)^2 - \frac{1}{2} (\partial \chi)^2 - V(\chi) \right]
\ee
The swampland conjecture for this theory reads,
\be
\frac{|V_{, \chi}|}{V(\chi)} \geq c ,
\ee
independent of $\phi$. Clearly, if the kinetic mixing between $\chi$ and $\phi$ induces dynamics for $\phi$, the above swampland conjecture does not capture the full dynamics of the theory. This potentially allows for interesting cosmological solutions without violating the above bound.

As discussed in \cite{Elder:2014fea}, provided the field $\chi$ has a positive mass, the two field model above is classically equivalent to a single-field higher-derivative action for $\phi$, of the form
\be
\label{PX}
\mathcal{L} = \sqrt{-g} P(X) \;\; , \;\; X =  - \frac{1}{2} (\partial \phi)^2  .
\ee
The mapping between the two descriptions is given by \cite{Elder:2014fea},
\be
f(\chi) = P_{,X} |_{X =M_{p}^3 \chi}  \;\; , \;\; V(\chi) = \left[ X P_{,X} -  P \right] |_{X = M_{p}^3 \chi}  ,
\ee
where $_{,X}$ denotes a derivative with respect to $X$. This matching is consistent provided $P_{,X} >0$. In general, $P(X)$ theories are valid even when $P(X)=0$, however in this case the matching is only classical and the $P(X)$ theory does not have a two field UV completion.

The utility of the higher derivative action \eqref{PX} is the ease of constructing de Sitter solutions. Such a solution exists at $X=X_0>0$ provided two conditions are satisfied:
\be
\label{PXdS}
P_{,X} (X_0) = 0  \;\; , \;\; P(X_0) = - \Lambda M_{p}^2
\ee
The first condition is required by the equations of motion, while the second condition is required for the solution to be de Sitter space. One can easily check that this indeed solves the equations of motion of the two-field model,
\bg
&&\ddot{\chi} + 3 H \dot{\chi} + f_{,\chi} X = V_{, \chi} \nonumber\\
&& f(\chi) \ddot{\phi} + 3 H f(\chi) \dot{\phi} + f_{,\chi} \dot{\chi} \dot{\phi}=0 ,
\nd
which are solved by $f(\chi)=P_{,X}=0$, $\dot{\chi}=0$, and $f_{,\chi} X - V_{, \chi} =0$. However, the requirement that $f(\chi)=0$ implies that the de Sitter solution does not have a two-field UV completion. Nonetheless, motivated by the connection to multifield models, we will proceed to study de Sitter solutions in $P(X)$ theories, and will use the multifield models only to rephrase the swampland conjecture.

Explicit examples of $P(X)$ models that give dS minima are not difficult to find. Consider the choice:
\be
P(X) = - \frac{X^2}{M_{p}^4} + \frac{X^4}{4 M_{p}^{10} \Lambda}  ,
\ee
where $\Lambda$, chosen for later convenience, has dimension of $M_{p}^2$.  Recalling the equation of motion is $\partial_t ( \sqrt{-g} \dot{\phi} P_{,X})=0$, this admits a solution satisfying $P_{,X}(X_0)=0$,
\be
X_ 0 = \sqrt{2} M_{p}^3 \sqrt{\Lambda} .
\ee
One can easily check that this satisfies the requirements for a dS solution \eqref{PXdS}. The solution for $\phi$ is time-dependent, and is given by
\be
\label{phit}
\phi(t) = \phi_0 + (8 \Lambda )^{1/4} M_{p}^{3/2} t 
\ee
Moreover, this solution satisfies the conditions for the quantum mechanical consistency of a $P(X)$ theory. The theory expanded about the $X=X_0$ vacuum is,
\be
P(X) = \frac{2  X^2}{M_{p}^4} + \frac{ X^4}{4 M_{p}^{10} \Lambda} +\frac{\sqrt{2} X^3}{\sqrt{\Lambda } M_{p}^7}-\Lambda  M_{p}^2, 
\ee
which satisfies the quantum mechanical consistency conditions of a $P(X)$ theory (see e.g.~(66) of \cite{Tsujikawa:2010sc}):
\be
P_{,X}  \geq 0 \;\; , \;\; P_{,X} + 2 X P_{, X X}  \geq 0   ,
\ee
where $X$ has been redefined as $X \rightarrow X + X_0$.

With the working $P(X)$ theory in hand, one can easily deduce the corresponding two-field model. For the above case, the  corresponding two field model is,
\be
\label{PXchif}
f(\chi) =  - 2 \frac{\chi}{M_{p}} + \frac{ \chi^3}{M_{p} \Lambda}  \;\; , \;\;
V(\chi) = - M_{p}^2 \chi^2 + \frac{3}{4} \left( \frac{M_{p}^2}{\Lambda} \right) \chi^4 . 
\ee
The $P(X)$ solution above corresponds to the time dependent $\phi$ given above and a constant VEV for $\chi$:
\be
\chi_0 = \frac{X}{M_{p}^3} =  \sqrt{ 2 \Lambda } .
\ee
This solution gives $f(\chi)=0$ and hence the two-field model cannot be consistently quantized in the dS minimum. Nonetheless, the mapping is consistent classically, and from this we can apply the dS swampland conjecture.

The conjecture takes a simple form, in Planck units $M_{p}=1$,
\be
\label{PXswampland}
\frac{|V_{, \chi}|}{V(\chi)}  = \frac{ |X_0 P_{,XX}(X_0)|}{\Lambda} > c ,
\ee
and thus becomes a constraint on the second derivative of $P(X)$. For the model above, the swampland conjecture reads,
\be
c ~ \lesssim ~ \frac{4 \sqrt{2}}{\sqrt{\Lambda}} ,
\ee
which is easily satisfied for a small cosmological constant. Indeed in the present universe, $\Lambda \sim 10^{-122} M_{p}^2$, and the bound reads $c < 10^{122}$. Thus we find that positive cosmological constant solutions can exist \emph{without violating the swampland conjecture}.  This also applies to quintessence or inflationary models that can be realized as a small deformation of this field configuration.

However, one can immediately see from \eqref{phit} that this $P(X)$ theory ceases to be an effective field theory at late times, $t ~ \sim ~ {1/ \sqrt{M_{p}\sqrt{\Lambda}}}$, 
 at which point the field excursion becomes Planckian $\Delta \phi \sim M_{p}$, violating the swampland distance conjecture. This points towards the possibility that while dS or quasi-dS solutions may exist without violating the swampland conjecture, these solutions are not low energy effective fields theories.

\section{Type IIB String Theory and stringy corrections to the Kahler Potential \label{example2.1}}

Motivated by the previous section, we now consider Type IIB string theory, and the effect of including perturbative corrections. As we will see, the leading $\alpha'$ correction introduces the necessary ingredients: modified kinetic terms and time-dependences, but in itself not sufficient to realize dS or violate the swampland conjecture. 


\subsection{No scale compactification}

We start by considering a no-scale compactification, specified by the Kahler potential
\be
\label{K0}
K_0 = -3 \log \left( T + \overline{T}\right) \equiv - 2 \log \mathcal{V},
\ee
and a constant superpotential:
\be
\label{W0}
W= W_0 .
\ee
The classical scalar potential of this theory is given by the standard $\mathcal{N}=1$ $d=4$ expression,
\be
V = |W_0|^2 e^{K} \left( K^{,T \bar{T} } K_{,T} K_{,\bar{T}} |W_0|^2 - 3 \right),
\ee
which vanishes for no-scale model. Considering only the single chiral superfield $T$, the mass spectrum is given by,
\be
\label{massesNoScale}
m_{\psi} ^2 = 3 m_{3/2}^2= \frac{3 |W_0|^2 }{\left( T + \overline{T}\right)^3}  \;\; , \;\;  m_{{\rm Re}T}^2 = 0 \;\; ,\;\;m_{{\rm Im}T}^2 = 0  ,
\ee
There is an additional tower of KK-modes, with masses quantized in units of
\be
m_{KK} =M_{p} \cdot \frac{M_{p}}{(T + \overline{T})}  .
\ee
When expressed in terms of the canonically normalized volume modulus 
${\varphi \over M_{p}}= \sqrt{3\over 2} \log\left({\phi\over M_{p}}\right)$, $\phi ={\rm Re}~ T$, the KK mass takes the form
\be
m_{KK} = M_{p} \cdot e^{- \sqrt{\frac{3}{2}} \varphi/M_{p}} ,
\ee
which realizes the swampland distance conjecture \eqref{dSSC2}.  Typical constructions of inflation in string theory take volumes ${\cal V}\sim 10^3$ and hence $\phi \sim 50 - 100 M_{p}$, in which case $m_{KK} \sim 10^{-2} M_{p}$ and the KK modes are very heavy. In particular, we note that this corresponds to super-Planckian values for the canonical scalar field $\varphi \sim 5 M_{p}$. 

Let us contrast the situation here with that of a $U(\chi)=m^2 \chi^2$ potential that is derived as a perturbation expansion about $\chi=0$. This is the case for complex structure moduli, which have a mass proportional to $|W_0|^2$, or for brane position moduli, with $\chi$ the fluctuation of the position of the brane about a fixed position $r_0$. In that case, one expects new light degrees of freedom for large field excursions, as discussed in \cite{vafa2}. However, for the volume modulus, \emph{the supergravity limit is only applicable in the limit that $\phi > M_{p}$ in the first place}. For smaller volumes, one expects massive string states to become massless. Thus it is only for intermediate volumes, e.g. $\phi \sim 100$, that this effective description is valid. Here we will show that the swampland conjecture cannot be violated in this regime.

As a final comment on the field range of $\phi$, we note that the initial condition for $\phi$ in the early universe may be in any of the three regimes: small, intermediate, large. The first case arises in string gas cosmology, while the second arises in string inflation, and the latter has been consider as a pre-inflationary state in \cite{Kontou:2017xhp}.

\subsection{String Theory corrections to Kahler Potential and the Swampland \label{example2.2}}

We now consider the effect of stringy corrections to the four-dimensional supergravity description which break the no-scale structure. As emphasized in \cite{Sethi:2017phn}, supersymmetry is broken by taking $W_0 \neq 0$, and in this in fact leads to a breakdown of perturbation theory, since the string-theoretic tower of no-scale breaking corrections to the K\"{a}hler potential will each generate a classical scalar potential for the resulting supergravity theory. One is forced to consider a corrected Kahler potential of the form,
\be
\label{Kexpanded}
K = K_0 + \sum_{nm}\delta K_{mn}\left(  {\alpha'}^n , g_s ^m \right) ,
\ee
where $\delta K_{mn}$ is the correction to the Kahler potential at order ${\alpha'}^n$ and $g_s ^m$.  

A systematic analysis of corrections to the K\"{a}hler potential was done in \cite{Berg:2005ja}. The result of \cite{Berg:2005ja} is summarized as \cite {Burgess:2010sy},
\be
K \simeq - 2 \log \, \mathcal{V} + \frac{k_1}{\mathcal{V}^{2/3}} + \frac{k_2}{\mathcal{V}} + \frac{k_3}{\mathcal{V}^{4/3}} + ...
\ee
where $k_1$, $k_2$, $k_3$ are parameters to be computed in a given model. The first term is a ${\alpha'}^2$ correction \cite{Pedro:2013qga, Grimm:2013gma}, the second term is the ${\alpha'}^3$ correction, and the third term is a string one-loop correction. Beyond this, the precise corrections are not known.

We will for the moment study only the leading $\alpha'$ correction, first worked out by \cite{Becker:2002nn}. The  Kahler potential is given by,
\bea K&=&-2\cdot \log\left({\cal
\hat{V}}+\alpha'^3\frac{\hat{\xi}}{2}\right), \nonumber \\
\hat{\xi}&=& -\frac{\zeta(3) \chi}{4\sqrt{2}(2\pi)^3}\;\cdot(S+\bar{S})^{3/2}\;\; ,
\eea 
 where $\chi$ is the Euler number of the Calabi-Yau, and $S$ is the dilaton. The dilaton $S$ is fixed at a constant value by the background fluxes,  encoded in $W_0$. One finds for the scalar potential,
 \be
 \label{Vxi}
V = - \frac{3 W_0 ^2 \xi}{\xi^3 - 24 \xi \phi^3 - 32 \sqrt{2} \phi^{9/2}} \simeq  + \frac{3 W_0^2 \xi}{32 \sqrt{2} \phi^{9/2}} + \mathcal{O}(\xi^2),
\ee
 while the mass formulae become
\bea
&& m_{3/2} ^2 = \frac{|W_0|^2}{8 \phi^3} \left( 1 - \frac{\xi}{\sqrt{2} \phi^{3/2}}\right)  \;\; ,\;\; m_{\psi} ^2 = 3 m_{3/2}^2 \\
&& m_{{\rm Re}T}^2 = \frac{297 |W_0|^2 }{128 \sqrt{2}} \frac{\xi}{\phi^{13/2}} \;\; ,\;\;m_{{\rm Im}T}^2 = 0  .
\eea 
This potential is of the runaway type: For positive $\xi$ (negative $\chi$) the volume runs away to $\infty$ while for negative $\xi$ (positive $\chi$) the volume collapses to $0$.  The potential for positive $\xi$ is shown in Figure 1.
\begin{figure}[h!]
\begin{center}
\includegraphics[scale=0.4]{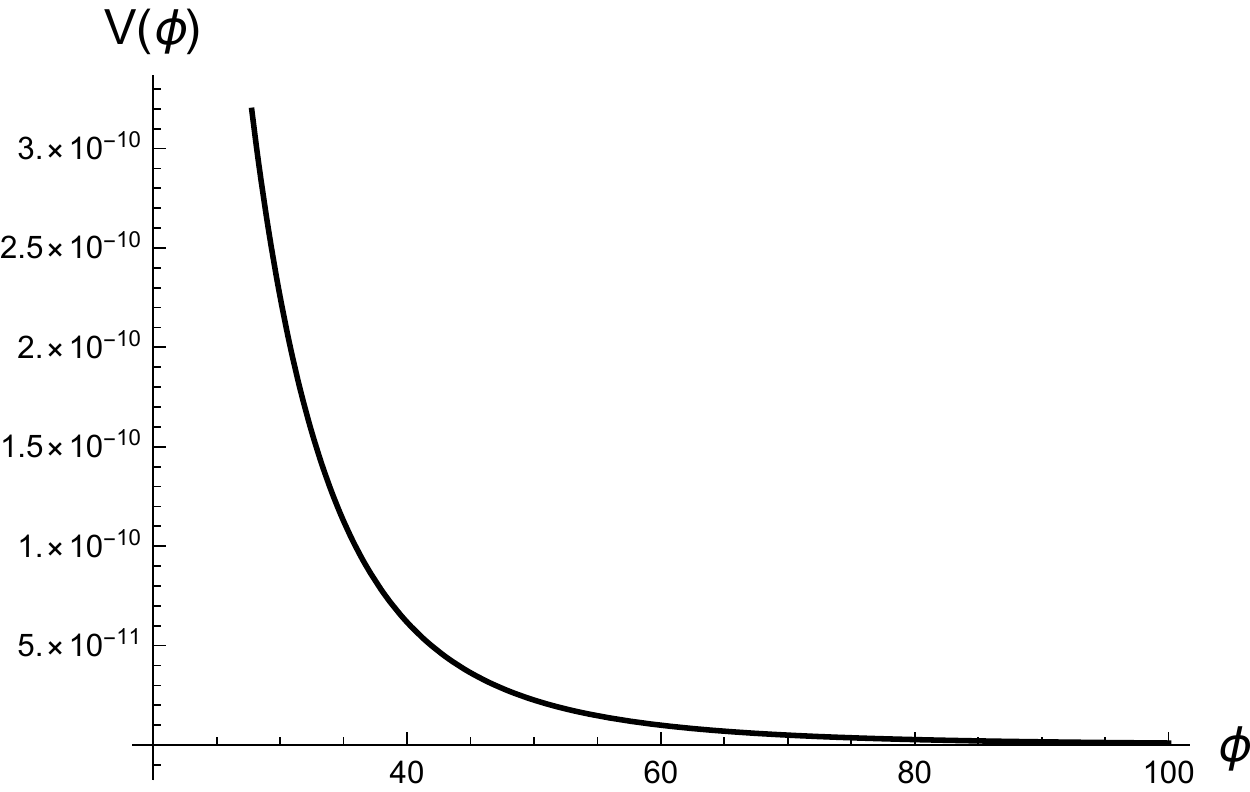}
\caption{Potential  $V(\phi)$ from leading $\alpha'$ correction with positive $\xi$ (i.e. negative $\chi$).}
\end{center}
\end{figure}

The lack of minimum at finite volume indicates that there is no stable background about which to compute non-perturbative corrections, as suggested in \cite{Sethi:2017phn}. However one may still study the perturbative dynamics in this unstable background, since despite the non-existence of a stable background solution, there do exists rolling solutions. 

With this in mind, we now turn to the swampland conjecture. Importantly, note that $\nabla V$ is defined as $\sqrt{g^{ij}\partial_{\phi^i} V \partial_{\phi ^j} V}$, where and hence depends on metric on the kinetic field space manifold. It is thus easiest to evaluate the swampland conjecture in terms of the canonical field,
\be
\frac{\varphi}{M_{p}}= \sqrt{\frac{3}{2}} \log \frac{\phi}{M_{p}} .
\ee
Then one can easily compute the quantities involved:
\be
\frac{|\nabla V|}{V} = \frac{9}{2 } \cdot  \sqrt{\frac{2}{3}}
\ee
And hence the swampland condition is satisfied for all values of $\phi$, even for $\phi \gg M_{p}$. However, similar to section \ref{sec:PX}, this setup violates the swampland distance conjecture at late times, indicating an eventual breakdown of effective field theory.

\section{Quantum Corrections and de Sitter from M-theory \label{mtheory}}

We now turn to a thorough analysis of the equations of motion. The swampland criterion, or the lower limit in \eqref{hurley}, rules out four-dimensional de Sitter solutions from string theory so the natural question is whether this is 
borne out of the no-go conditions proposed in \cite{nogo} and \cite{hirano}. Both of these no-go conditions are in fact the refined forms of the no-go conditions originally proposed in \cite{gibbons} and \cite{malnun}, and deal with eliminating all classical sources, including orientifold planes and anti-branes, that were originally thought of giving rise to four-dimensional de Sitter vacua from string theory. In \cite{nogo} it was proposed that a severe fine-tuning may be required to realize any hope of getting a four-dimensional de Sitter solution.  Whether such fine-tunings are indeed possible was not discussed in \cite{nogo}, and here we want to not only discuss this aspect of the construction but also measure it against the swampland criteria of \cite{vafa1, vafa2}.

\subsection{Vacua with de Sitter isometries and their M-theory uplifts}

Our starting point then is the assumption that there does exist a four-dimensional de Sitter solution in type IIB string theory, and we ask what kind of quantum corrections are required to fully realize such a solution. To simplify certain aspects of the computations, we will use M-theory uplift to study the various ingredients entering our analysis. Needless to say such uplifting do not change any physical aspects of the results, and we could have analyzed this directly from type IIB also as was shown in \cite{hirano}, but the sheer brevity of the expressions from M-theory is an attractive alternative to the somewhat tedious exercise of keeping track of the multiple fields in type IIB theory. 

The type IIB background that we want can be expressed in terms of a six-dimensional compact internal space with an unwarped metric $g_{mn}$ in the following way:
\bg\label{li2018}
ds^2 =   {1\over \Lambda(t) \sqrt{h}} \left(-dt^2 + dx_1^2 + dx_2^2 + dx_3^2\right) + \sqrt{h} g_{mn} dy^m dy^n, \nd
where $\Lambda(t)$ is a time-dependent function, $y^m$ are the coordinates of the internal space, and $h$ is the warp-factor that could be a function of all the internal coordinates. For simplicity however we can take $h(r)$ to be a function of the internal radial coordinate. The dilaton remains a constant so the string and the Einstein metric coincides.
Note that the way we represented the background, the internal space is time {\it independent} and therefore the four-dimensional Newton's constant will be time independent.  We will also take:
\bg\label{bava}
\Lambda(t) \equiv \Lambda \vert t\vert^2, \nd
with $\Lambda > 0$ so that the four-dimensional metric in \eqref{li2018} is indeed a de Sitter space\footnote{We are essentially using the so-called {\it flat slicing} for a four-dimensional de Sitter space. This does not globally cover de Sitter, in fact they only cover the top triangle, although there is a relation between flat slicing and global coordinates. Note that in this coordinate system time flows from $t = -\infty$ to $t = 0$.}. 
Our aim now is to answer two set of questions that are related to \eqref{li2018}. The first is to ask whether all the fluxes and the quantum corrections that go in the construction of \eqref{li2018} can be time independent. The second is to investigate on the  {\it minimal} set of quantum corrections needed to actually realize a background of the sort given in \eqref{li2018}.

Answering both these questions will take us to M-theory where the analysis will be much more tractable, as mentioned earlier. To lift the background \eqref{li2018} to M-theory, we will compactify the $x_3$ direction to a circle and fiber it over the $x_{11}$ circle to form a torus ${\bf T}^2$ of complex structure $\tau = i$. Let us denote the complex coordinate of the torus by $z$.  As is well known, in M-theory we have to deal with a eight-dimensional manifold,  which is in fact  the torus ${\bf T}^2$ fibered over a six-dimensional base. However compared to what we had in 
\eqref{li2018}, neither the base nor the fiber of our eight-dimensional space can be time independent. The precise 
metric is \cite{nogo}:
\bg\label{sandman}
ds_8^2 = {g_{mn}dy^mdy^n\over \Lambda^{1/3} \vert t\vert^{2/3}} + \Lambda^{2/3} \vert t\vert^{4/3} \vert dz\vert^2, \nd
where one could see that as time progresses the size of the six-dimensional base increases whereas the size of the fiber torus shrinks. This will take us to type IIB theory at late times but the background there may not have the full de Sitter isometries. In the M-theory side, one might be worried that supergravity analysis could not be valid at early times if the size of the base approaches Planck's length, and it is indeed a genuine concern so for the time being we will restrict our study within the allowed range of time evolution. Of course as we shall soon see, quantum effects are essential to consistently realize a background of the form \eqref{li2018}, so in the end it is not just supergravity analysis that we require, but a more detailed  analysis with quantum corrections. On the other hand in the type IIB side, as is evident from \eqref{li2018}, nothing untoward happens to the internal six-dimensions at early times.  

Keeping all these in mind, let us first switch on G-fluxes $\mathbb{G}_{mnpq}, 
\mathbb{G}_{mnpa}$ and $\mathbb{G}_{mnab}$ on the internal eight-dimensional space, where $m, n$ denote coordinates on the six-dimensional base, and $a, b$ denote coordinates on the fiber torus. These fluxes generically cannot be all time independent, but we can keep the components $\mathbb{G}_{mnpa}$ to be time-independent without loss of generalities. However, as argued in 
\cite{nogo}, to solve EOMs consistently we also require a spacetime component of the three-form flux of the form:
\bg\label{cold}
 C_{\mu\nu\rho} ~ = ~ {\epsilon_{\mu\nu\rho}\over h \Lambda^2 \vert t\vert^4}, \nd
 where $\mu, \nu$ denote 2+1 dimensional spacetime coordinates and $h$ is the warp-factor appearing in 
 \eqref{li2018}. The three-form flux increases in value as time progresses, and the time dependence of \eqref{cold} and some of the other G-flux components already tell us that the background 
 \eqref{li2018} when dimensionally reduced to four-dimensions will have time dependent moduli fields therefore cannot quite have all the de Sitter isometries. Such dependences should remind the readers of \eqref{vidya} discussed earlier, but at this stage it is a bit premature to make concrete conclusions. 
We need more details, and in the following we elaborate the story further emphasizing on the EOMs and the quantum effects. 
 
The first issue that we want to concentrate on is the time dependence of the internal fluxes exemplified above. For concreteness we have taken all possible components of the fluxes, and using these let us define an integral of the form:
\bg\label{tagradesi}
\mathbb{I}_1 = \int_{\Sigma^8} d^8x \sqrt{{\bf g}_8} \left(\mathbb{G}_{mnpq}\mathbb{G}^{mnpq} + \mathbb{G}_{mnpa}\mathbb{G}^{mnpa} + \mathbb{G}_{mnab}\mathbb{G}^{mnab}\right), \nd
subset of which is basically the terms responsible for a part of the cosmological constant. In type IIB language the first term is the five-form contribution, the second term is the  NS and RR three-form contributions and the third terms is the RR three-form contribution.  In other words we are switching on:
\bg\label{icehouse}
&& \mathbb{C}_{mna}(y, t) = C_{mna} (y) + \delta C_{mna}(y, t) \nonumber\\
&& \mathbb{C}_{mnp}(y, t) = C_{mnp}(y, t) + \delta C_{mnp}(y, t) \nonumber\\
&& \mathbb{C}_{mn3}(x_{11}, t) = C_{mn3}(x_{11}, t) + \delta C_{mn3}(x_{11}, t), \nd
where $\delta C$ denote fluctuations above the background values of the corresponding fields. Note that, according to our choice, we have kept $C_{mna}$ to be time-independent.  However subtleties appear from the time dependences of the background values of other two components in \eqref{icehouse} as de Sitter isometries can be broken from their time-dependences. One way out would be to take vanishing values for the ${G}_{mnpq}$ and 
${G}_{mnab}$ components. Alternatively, we can assume that the time-dependences are very slow\footnote{In other words, $ \partial_{[m} C_{npq]} \gg {\dot C}_{npq}$. \label{slowt}}. Taking the latter into considerations, 
the above choices of the internal fields now give rise to another integral of the form:
\bg\label{hardivine}
\mathbb{I}_2 = \int_{\Sigma^8} d^8x \sqrt{{\bf g}_8} \left(\mathbb{G}_{0npq}\mathbb{G}^{0npq} + \mathbb{G}_{0npa}\mathbb{G}^{0npa}\right), \nd
where the absence of the $\mathbb{G}_{0nab}$ piece can be accounted from the choice \eqref{icehouse}.  The logic behind the two integrals, \eqref{tagradesi} and \eqref{hardivine}, should be clear from the following decomposition of the three-form flux in M-theory:
\bg\label{ctrust}
{\delta C}_{MNP}(y, t) = \sum_{i = 1}^{b_3} \varphi^i(y, t) \otimes \Omega^{(i)}_{MNP}(y), \nd
where $\Omega^{(i)}_{MNP}(y)$ are the harmonic three-forms on the eight-manifold $\Sigma_8$ in \eqref{sandman} and $b_3$ is the third Betti number of $\Sigma_8$. The next level of subtlety now appears from the time dependence in \eqref{sandman}. However if the topology of $\Sigma_8$ does not change, we expect $\Omega^{(i)}_{MNP}$ to not change 
very much and so all variations with respect to time should appear in the scalar fields $\varphi_i$. Such a state of affair then tells us that $\mathbb{I}_2$ and $\mathbb{I}_1$ precisely give us the kinetic and the potential terms of the scalar fields respectively.  In fact if we also allow the other components of the G-flux namely 
$\mathbb{G}_{\mu mnp}$ and $\mathbb{G}_{\mu mna}$, where $\mu$ are all the three-dimensional coordinates, then:
\bg\label{chukkamukha} 
\int d^3 x \sqrt{-{\bf g}_3} \left(\mathbb{I}_1 + \mathbb{I}_2\right) = 
- \int d^3 x \sqrt{-{\bf g}_3}\left[{1\over 2} \sum_{k = 1}^{n} \partial_\mu \phi_k \partial^\mu \phi_k 
- V\left(\{\phi_k\}\right)\right], \nd
with $\phi_k$ now denoting the canonically normalized scalars from $\varphi_k$, and $V\left(\{\phi_k\}\right)$ being the potential of all the scalars appearing in the spectrum. Note that the upper limit of the sum in \eqref{chukkamukha} 
is no longer $b_3$ but $n \equiv b_3 + h_{11} + 2h_{31}$ that involves the Hodge numbers $h_{11}$ and $h_{31}$ of the eight-manifold\footnote{Assuming of course that the eight-manifold $\Sigma_8$, at any given time, allows for an integrable complex structure. If it doesn't then the analysis will be more involved. Here we will avoid these subtleties.}. These additional scalars come from the metric fluctuations. 
 
The story now is closer to what we discussed in section \ref{swampland}, albeit now in three-dimensions. The M-theory analysis reproduces a potential $V\left(\{\phi_k\}\right)$ for the scalars $\phi_k$ in the dimensionally reduced spacetime. For all of these to make sense the potential has to be time independent as in \eqref{jlilharmony}.  
The time derivative of $\mathbb{I}_1$ gives us:
\bg\label{goeskalu}
{\partial\mathbb{I}_1\over \partial t} & = & \int_{\Sigma^8} d^8 x {\sqrt{{g}_8}\over h\Lambda \vert t\vert^2} 
\left[{\partial\over \partial t} \left(G_{mnpq}G^{mnpq}\right) + {1\over \Lambda^2 \vert t\vert^4} {\partial\over \partial t} 
\left(G_{mnab}G^{mnab}\right)\right] \nonumber\\
& - & \int_{\Sigma^8} d^8 x {2 \sqrt{{g}_8}\over h\Lambda \vert t\vert^3}\left(G_{mnpq}G^{mnpq} + 
{3G_{mnab}G^{mnab}\over \Lambda^2 \vert t\vert^4} + {2G_{mnpa}G^{mnpa}\over \Lambda \vert t\vert^2}\right),
\nd
where $G_{mnpq}, G_{mnab}$ and $G_{mnpa}$ are the unwarped versions of the G-fluxes as given in eq. (5.9) of 
\cite{nogo}. Similarly $g_8$ is the unwarped metric of the eight-dimensional space $\Sigma_8$ given in 
\eqref{sandman}. A similar integral, but with slightly different time-dependent coefficients, will appear in the type IIB side too but because of the self-dual nature of the five-form it is difficult to write this explicitly. However one might go to the non self-dual type IIB action where an integral like \eqref{goeskalu}  can indeed be written (with additional constraints). 

Demanding the vanishing of such an integral seems to imply time-dependences of the G-flux components, although in the type IIB side if we only retain the $\mathbb{G}_{mnpa}$ components, which are time-independent, the integral 
\eqref{goeskalu} could vanish\footnote{This in turn will also avoid introducing non self-dual type IIB action.}.
Of course we haven't yet introduced any quantum effects so any such conclusions based on classical flux configurations should be considered incomplete. In the following section, we therefore proceed to the next stage of our analysis related to the quantum corrections. 
 
 \subsection{Equations of motion and quantum effects  from M-theory \label{EOMs}}

The quantum corrections appear from variety of sources in M-theory: there are higher order curvature corrections, higher order G-flux corrections, and M2 and M5-brane instantons corrections.  As discussed in \cite{nogo}, these corrections may be divided into topological and non-topological pieces and in the following we will concentrate 
{\it only} on the polynomial parts of the bosonic non-topological corrections around weak curvatures and small G-flux field strengths. These may be expressed as\footnote{In string theory or M-theory the fields are taken to be dimensionless, therefore it is the derivative expansion that matters. This is used to fix the dimensions.}:
\bg\label{gaiman}
V_{\mathbb{Q}} \equiv  \sum_{m, n, p, q} \int d^8x \sqrt{{\bf g}_8}\left({{\cal C}_{mnpq} \square^m {\bf R}^n \square^p {\bf G}^q + {\cal D}_{mnpq} \square^m \left({\bf R}^n\right)_{rs} \square^p \left({\bf G}^q\right)^{rs} + ....\over M_{p}^{2m + 2p + q + 2n- 8}}\right), \nd
where ${\cal C}_{mnpq}, {\cal D}_{mnpq}$ are constants and the powers of the curvature and the G-flux, which are raised and lowered by the warped metric components, are contracted in appropriate ways as described in \cite{deser, grosswitten, pisin}. The dotted terms involve higher tensorial contractions. All these terms are suppressed by powers of $M_{p}$, and we will discuss later what kind of hierarchies exist between them. For example ${\bf R}^n$ and ${\bf G}^m$  may be constructed using multiple  possible 
contractions\footnote{For example ${\bf R}^2 \equiv c_0 R^2 + c_1 R_{mn}R^{mn} + c_2 R_{mnpq} R^{mnpq} + ....$, where $c_i$ are constants and the dotted terms are other possible contractions. Similar expansions may be made for ${\bf G}^2$ and higher powers of $R$ and $G$. In fact existence of terms like these can help us to get the Born-Infeld action with multi Taub-NUT spaces for type IIA D6-branes \cite{sav1, moore}.}.  
Thus to make sense of $V_{\mathbb{Q}}$ one will have to impose some extra hierarchies to restrict the series to only finite number of terms.  For the time being we will proceed without worrying too much about this, and express the quantum pieces as the following three-dimensional integral:
\bg\label{DOA}
\mathbb{I}_3 =  M_p^3\int d^3 x \sqrt{-{\bf g}_3} V_{\mathbb{Q}}. \nd
Combining this with \eqref{chukkamukha} will give us the complete picture in three-dimensions. If we shrink the 
fiber-torus of $\Sigma_8$ to zero size this will take us to type IIB theory where we can have a four-dimensional description. The quantum piece therein will be related to \eqref{gaiman} via the duality map, and so would be the energy-momentum tensor whose typical form, in three-dimensions, may be written as \cite{nogo}:
\bg\label{20list18} 
T_{MN} = -{2\over \sqrt{-{\bf g}_{11}}} {\delta \mathbb{I}_3 \over \delta {\bf g}^{MN}} \equiv 
\sum_i h^{1/3}\left(\Lambda\vert t \vert^2 \right)^{\alpha_i} \mathbb{C}^{(i)}_{MN}, \nd
where $\mathbb{C}_{MN}^{(i)}$ are functions of $h, g_{MN}, M_{p}$ and the background three-form flux $C_{MNP}$. The time dependences are extracted out of each terms in such a way that $\mathbb{C}_{MN}^{(i)}$'s are all time independent. Clearly if we arrange the series \eqref{20list18} with {\it decreasing} $\alpha_i$, i.e if we make the following arrangements:
\bg\label{ericajong}
\alpha_i > \alpha_{i+1}, \nd
then there is some hierarchy between the various quantum terms, at least perturbatively. 
This hierarchy will be lost if $\alpha_i = 0$. The logic behind such an arrangement is to note that the type IIA coupling $g_s$ is proportional to:
\bg\label{cashjul}
g_s  ~ \propto ~ \left(\Lambda\vert t\vert^2\right)^{1/2} h^{1/4}, \nd
which decreases slowly with time towards weak coupling, but is strongly coupled at early times. In this sense M-theory is the correct description of the background at early times, and therefore \eqref{ericajong} does indeed provide some hierarchy between the various quantum pieces when $g_s < 1$.\footnote{Note however that the situation at hand is more subtle than one might have anticipated. Consider for example the case where $g_s  < 1$ at certain time $t_0$. The hierarchy that we gain with such 
weak-coupling scenario quickly fades away when $g_s  > 1$ at an early time. The effective field theory is no longer under control at an earlier time as an infinite series of higher order corrections become relevant. This is of course one of the many issue that one would face describing a four-dimensional theory from the type IIA side, but the breakdown of any effective field theory that we want to discuss here will be unrelated to the early time strong coupling effect that appears here. One interesting point to note however is that, as $t \to -\infty$, the three-form flux \eqref{cold} goes to zero faster than the M-theory metric. If the other flux components are time dependent, and goes to zero at early times, then this is where the full de Sitter isometries should be visible in the type IIB side.} To see how this works out precisely let us first consider some time-neutral terms in the quantum sum:
\bg\label{dqueen} 
&& \Lambda_{(1)} \equiv {{\bf G}^{mnpq} {\bf G}_{mn}^{~~~ab} {\bf G}_{abpq}\over M_{p}^3}, ~~
\Lambda_{(2)} \equiv {{\bf R}^2 {\bf R}^{ab} {\bf R}_{ab} {\bf G}^{mnab}{\bf G}_{pqab}{\bf G}^{pqcd} {\bf G}_{cdmn}
\over M_{p}^{12}}\\
&& \Lambda_{(3)} \equiv {{\bf G}_{rsab}{\bf G}^{rsab}{\bf R}_{[mn][pq]} {\bf R}^{[mn]}{\bf G}^{pqcd} {\bf R}_{[cd]} \over M_{p}^9}, ~~
\Lambda_{(4)} \equiv {{\bf R} {\bf R}_{mnpq} {\bf G}^{mn}_{~~~ab} {\bf G}^{pqab}\over M_{p}^6}\nonumber\\
&& \Lambda_{(5)} \equiv {{\bf R}^2 {\bf G}_{mnab} {\bf G}^{mnab} \over M_{p}^6}, ~~ 
\Lambda_{(6)} \equiv {{\square}^2 {\bf G}_{mnab} {\bf G}^{mnab}\over M_{p}^6}, ~~
\Lambda_{(7)} \equiv  {\left({\square} {\bf R}\right) {\bf G}_{mnab} {\bf G}^{mnab}\over M_{p}^6}, \nonumber \nd
where we have assumed that the warp-factor $h$ is a function of all the coordinates of the eight-manifold so that 
${\bf R}_{ab}$ is non-zero, the G-flux components with all lower indices are time independent\footnote{The fact that this is indeed possible is the subject of this section and will be rigorously demonstrated below.}, 
and $\square$ here is defined with respect to the six-dimensional base 
only\footnote{$\square \equiv {1\over \sqrt{{\bf g}_6}} \partial_m \left(\sqrt{{\bf g}_6} {\bf g}^{mn} \partial_n\right)$ where ${\bf g}_6$ is the determinant of the six-dimensional base metric of $\Sigma_8$.}. Note that $\Lambda_{(i)}$'s do not scale with respect to time but appear with different orders in $M_{p}$. Now imagine we want to extract a term of the form $\left(\Lambda\vert t\vert^2\right)^{\alpha_i}$ from the quantum series \eqref{DOA}. Such a term can appear from various pieces in the quantum series \eqref{DOA} or \eqref{gaiman} in \eqref{20list18}. For example let us first construct the following series:

{\footnotesize
\bg\label{dasecret1}
 && {1\over M_{p}^{6a_\gamma}}\left(\sum_{\{n_k\}} \mathbb{C}_{\{n_k\}} \prod_k \Lambda^{n_k}_{(k)} \right)
 \left[{{\bf R}^{3a_\gamma} {\bf g}_{n}^{~w} \over M_{p}^{0}} + 
 {{\bf R}^{3a_\gamma - 4} {\bf G}^w_{~~pqr} {\bf G}_n^{~~pqr} \over c_1 M_{p}^{ - 6}} + 
 {{\bf R}^2 {\bf G}^w_{~~pab} {\bf G}_n^{~~pab} \left({\bf R} {\bf R}_{cd} {\bf R}^{cd}\right)^{a_\gamma}\over c_2 
 M_{p}^{8}} + ..... \right]
 \nonumber\\
 && = ~  \left(\Lambda\vert t\vert^2\right)^{\vert a_\gamma\vert} \mathbb{D}_{n}^{(\gamma)w} 
  \nd}
where the terms in the first bracket above are all Lorentz scalars with integer $\mathbb{C}_{\{n_k\}}$, $c_i$ are dimensionless constants, and 
the appearance of negative powers of $M_p$ is because of our choice of large $a_\gamma$. 
To identify \eqref{dasecret1} to the full $\mathbb{C}_{MN}^{(\gamma)}$ 
in \eqref{20list18} one will have to work out another series by
multiplying the series in the first bracket  of \eqref{dasecret1} to yet another time-neutral rank two tensor series constructed like \eqref{dqueen} in the following way: 

{\footnotesize
\bg\label{dasecret2}
 && {1\over M_{p}^{12a_\eta}}\left(\sum_{\{n_l\}} \mathbb{C}_{\{n_l\}} \prod_l \Lambda^{n_l}_{(l)} \right)
\left[{\left({\bf R} {\bf G}_{pqcd} {\bf G}^{pqcd}\right)^{3a_\eta - 3} {\bf G}_m^{~~rab} {\bf G}_{wrab}\over M_p^{-10}} +
{\left({\bf R} \square^{'3}{\bf R}_{[ab]}\partial_{[q}{\bf R}\partial_{n]}{\bf G}^{qnab}\right)^{a_\eta} {\bf R}_{(mw)}\over 
d_1 M_p^{3a_\eta + 2}} + ....\right] \nonumber\\
&& = ~\left(\Lambda\vert t\vert^2\right)^{-\vert a_\eta\vert} \mathbb{E}_{mw}^{(\eta)}, 
\nd} 
where $\square'$ is defined as the Laplacian along the torus direction of $\Sigma_8$, $d_i$ are dimensionless constants, and 
${\bf R}_{[MN]}$ and ${\bf R}_{(MN)}$ denote the anti-symmetric and the symmetric parts of ${\bf R}_{MN}$. Note that
the power of $\Lambda\vert t\vert^2$ is now a negative integer $-\vert a_\eta\vert$, compared to what we had in \eqref{dasecret1}. This implies that we can multiply the two series \eqref{dasecret1} and \eqref{dasecret2} 
to get the following new series:
\bg\label{dasecret3}
 \mathbb{C}_{mn}^{(\eta,\gamma)} & = &\sum_w  \mathbb{E}_{mw}^{(\eta)}\mathbb{D}_{n}^{(\gamma) w} 
=   \sum_w {\left(\Lambda\vert t\vert^2\right)^{\vert a_\eta\vert - \vert a_\gamma\vert}\over M_p^{12a_\eta + 6 a_\gamma}}
\left(\sum_{\{n_l\}} \mathbb{C}_{\{n_l\}} \prod_l \Lambda^{n_l}_{(l)} \right)\left(\sum_{\{n_k\}} \mathbb{C}_{\{n_k\}} \prod_k \Lambda^{n_k}_{(k)} \right)\\
&\times & \Bigg[{\left({\bf R} {\bf G}_{pqcd} {\bf G}^{pqcd}\right)^{3a_\eta - 3} {\bf G}_m^{~~rab} {\bf G}_{wrab}\over M_p^{-10}} + ......\Bigg]\Bigg[{{\bf R}^2{\bf G}^w_{~~u\sigma\delta} {\bf G}_n^{~~u\sigma\delta} \left({\bf R} {\bf R}_{\alpha\beta} {\bf R}^{\alpha\beta}\right)^{a_\gamma}\over c_2 M_{p}^{8}}+ ......\Bigg].  \nonumber  \nd
The above form of the expression is more useful than any of the two series \eqref{dasecret1} and \eqref{dasecret2} because it not only tells us how to extract any powers of $\Lambda\vert t\vert^2$, but also gives us a way to rewrite the time-neutral series in a more elegant way.  As a {\it first} trial, let us identify $\mathbb{C}_{mn}^{(k)}$ of \eqref{20list18} 
with $\mathbb{C}_{mn}^{(\eta,\gamma)}$ in the following way:
\bg\label{oshotmeye}
\mathbb{C}_{mn}^{(k)} ~ \equiv ~M_p^{2} 
 \mathbb{C}_{mn}^{(\eta, \gamma)}, ~~~~~
 k \equiv \vert a_\eta\vert - \vert a_\gamma \vert, \nd
implying that we need to scan  a range of two integers for a given choice of the integer $k$. The above approach points out to an {\it infinite} degeneracies for any given value of $k$. For example,  
the set of zeroes of $k$ includes an infinite range of integers of the form:
\bg\label{hgtko}  
\Big\{ k = 0 ~\Big\vert \left(a_n, a_n\right) ~ \forall ~n\Big\},  \nd
and similarly for other choices of $k$. Clearly such infinite degeneracies are not visible from either of the two parent series \eqref{dasecret1} and \eqref{dasecret2}, but comes out in the open once we follow the above procedures. Therefore, after the dust settles, the $\mathbb{C}_{mn}^{(k)}$ functions may be succinctly presented as the following series:
\bg\label{dasecret}
\mathbb{C}_{mn}^{(k)} & = &   \sum_w {\left(\Lambda\vert t\vert^2\right)^k \over M_p^{12(k + \vert a_\gamma\vert) 
+ 6 \vert a_\gamma\vert - 2}}
\left(\sum_{\{n_l\}} \mathbb{C}_{\{n_l\}} \prod_l \Lambda^{n_l}_{(l)} \right)\left(\sum_{\{n_m\}} \mathbb{C}_{\{n_m\}} \prod_m \Lambda^{n_m}_{(m)} \right)\\
&\times & \Bigg[{\left({\bf R} {\bf G}_{pqcd} {\bf G}^{pqcd}\right)^{3(k + \vert a_\gamma\vert) - 3} {\bf G}_m^{~~rab} {\bf G}_{wrab}\over M_p^{-10}} + ....\Bigg]\Bigg[{{\bf R}^2{\bf G}^w_{~~u\sigma\delta} {\bf G}_n^{~~u\sigma\delta} \left({\bf R} {\bf R}_{\alpha\beta} {\bf R}^{\alpha\beta}\right)^{\vert a_\gamma\vert}\over c_2 M_{p}^{8}}+ ...\Bigg].  \nonumber  \nd
for all integer values of $a_\gamma$ giving rise to the infinite degeneracies. Thus for a given value of $k$, what linear combinations are actually chosen can only be determined once we know the full quantum expansion, i.e all the terms in \eqref{gaiman}, of M-theory.  Note that $\left(\Lambda\vert t\vert^2\right)^{\vert k\vert}$ in \eqref{dasecret} may be easily constructed from \eqref{dasecret1} by contracting the series with ${\bf g}^n_w$, thus forming a Lorentz invariant series. Similarly $\left(\Lambda\vert t\vert^2\right)^{-\vert k\vert}$ may  be easily constructed from \eqref{dasecret2} 
by contracting the series \eqref{dasecret2} by  ${\bf R}^2 {\bf g}^{mw}$ 
and identifying the exponent to $-\vert k\vert$ accordingly\footnote{For large exponent, another choice includes contracting the series \eqref{dasecret2} with $\left({\bf G}_{pqab}{\bf G}^{pqab}\right)^2 {\bf g}^{mw}$ to form a Lorentz invariant series. Similar story goes for positive exponent.}. This way \eqref{dasecret} can be represented completely as a rank 2 symmetric tensor constructed out of ${\bf R}$, ${\bf G}$ and their derivatives.
At this stage, we can even relax the symmetric property of the tensor to allow for inherent torsion in the background, although note that \eqref{dasecret} is not the most generic answer we can have for $\mathbb{C}_{mn}^{(k)}$. We can combine six different series, like \eqref{dasecret1} and \eqref{dasecret2}, and contract them in the standard way to construct the $\mathbb{C}_{mn}^{(k)}$ series. In a similar vein the other two series for $\mathbb{C}^{(k)}_{ab}$ and $\mathbb{C}^{(k)}_{\mu\nu}$ can also be constructed. Using these, the energy-momentum tensor $T_{MN}$ from \eqref{20list18}
may now be rewritten, using \eqref{dasecret}, in the following way:
\bg\label{rakth}
T_{MN} = \sum_k h^{1/3}\left(\Lambda\vert t \vert^2 \right)^{\alpha_k} \mathbb{C}^{(k)}_{MN} = 
\sum_k \left(\Lambda\vert t\vert^2\right)^{\alpha_k + k} \mathbb{J}^{(k)}_{MN}, \nd
where the functional form for $\mathbb{J}_{mn}^{(k)}$ can be easily extracted from \eqref{dasecret} and from the equivalent series for $\mathbb{C}^{(k)}_{ab}$ and $\mathbb{C}^{(k)}_{\mu\nu}$. The above arrangement is a useful way to organize the series, but is not necessarily unique. Other arrangements are clearly possible, and we will discuss them later. Either way, from \eqref{rakth} we see that $T_{MN}$ is in general a function of time, and the time-independent contributions come from the following set:
\bg\label{mdtension}
\{ \alpha_k \}  =  0, ~~~~ \forall ~k, \nd
which are precisely the infinitely degenerate functions \eqref{dasecret}.  
Additionally,  
from \eqref{dasecret1}, \eqref{dasecret2} and \eqref{dasecret} we see that to the same order in $g_s$ the terms in the brackets appear with various powers of $M_{p}$. Although the $g_s$ expansion is creating a specific hierarchy compared to the $M_{p}$ expansion, and is seemingly better suited at arranging the quantum corrections, the  convergence\footnote{By {\it convergent} we will henceforth mean, unless mentioned otherwise, well-behaved or controlled.} of the series for a given power of $g_s$  is in question now. From the preliminary analysis presented here, it is not guaranteed the series will be convergent\footnote{In \eqref{dasecret}, for example,  there are in fact {\it three} different series running in parallel. One, is with respect to the time-neutral and Lorentz invariant functions, some of which are being collected in 
\eqref{dqueen}. Two, is with respect to the factor of $\left(\Lambda\vert t\vert^2\right)^k$ which may be expressed 
by appropriately constructing Lorentz invariant series from \eqref{dasecret1} and \eqref{dasecret2}, 
and finally three, is with respect to the symmetric rank two tensor functions for a given power of 
$\Lambda\vert t\vert^2$. The convergence properties of all these series are not guaranteed, at least from the simple analysis that we presented here.}.  
Despite this, at strong coupling it will still be the right expansion parameter because the equations of motion, that we will analyze below, will again be better expressed in powers of $\Lambda\vert t\vert^2$. This way matching the quantum pieces can be performed efficiently.   


This now brings us to the point where we will have to analyze all the EOMs of the system carefully. Fortunately some aspect of this is already studied in \cite{nogo} and \cite{pisin} so our work will be somewhat simplified. However compared to
\cite{nogo} and \cite{pisin}, we will keep track of $G_{mnp0}$ piece just for completeness sake. 

The metric of the six-dimensional base $g_{mn}$ of the eight-manifold $\Sigma_8$, can be determined by the time-independent flux $G_{mnpa}$ and the quantum corrections in the following way:
\bg\label{megrya}
G_{mn} - 6\Lambda h g_{mn} = {1\over 4h}\left(G_{mpqa}G_{n}^{~~pqa} - {1\over 6} g_{mn} G_{pqra}G^{pqra}\right) +
h^{1/3}\sum_{\{\alpha_i\} = 0}\mathbb{C}^{(i)}_{mn}, \nd
where $G_{mn}$ is the Einstein tensor constructed out of the unwarped metric $g_{mn}$. Note that the quantum pieces appear as {\it sum} over all the set of $\alpha_i$ that vanish i.e sum over all the set in \eqref{mdtension}. For a given value of $\alpha_i$, we already raised the issue of  convergence in \eqref{dasecret}. Now that we are dealing with the {\it sum} of all the 
$\mathbb{C}_{mn}^{(i)}$ functions, the situation is more acute now. Even if we assume that each of the series in 
\eqref{dasecret}, for a given set of $\alpha_i$, is convergent,  
unless this set is of finite size there isn't much hierarchy between the various quantum pieces. We will discuss more on this in section \ref{nala}. The time-dependent fluxes, on the other hand, are related via the following EOM:
\bg\label{rampling}
&& {\Lambda \vert t \vert^2 \over 12 h}\left(G_{mpqr} G_{n}^{~~pqr} - {1\over 8} g_{mn} G_{pqrl}G^{pqrl}\right) 
+ {1\over 4h \Lambda \vert t \vert^2}\left(G_{mpab} G_{n}^{~~pab} - {1\over 4} g_{mn} G_{pqab}G^{pqab}\right)\nonumber\\
&& +  {\Lambda^2 \vert t \vert^4 \over 4}\left(G_{mpq0} G_{n}^{~~pq0} - {1\over 6} g_{mn} G_{pqr0}G^{pqr0}\right) 
+ h^{1/3} \sum_{\{\alpha_i\} \ne 0}\left(\Lambda \vert t \vert^2\right)^{\alpha_i} \mathbb{C}^{(i)}_{mn} = 0, \nd  
where the quantum pieces are defined by the set of all $\alpha_i$ that do not vanish and because of our choice 
\eqref{icehouse} we do not have any $G_{0mab}$ pieces in \eqref{rampling}. However the time-dependent fluxes, as mentioned earlier, break the de Sitter isometries so with these fluxes the background in type IIB side cannot be strictly de Sitter, although it will be a background with positive cosmological constant. A way out, suggested earlier,  would be to take the flux components to be very slowly varying with time and, following footnote \ref{slowt}, ignore the 
$G_{0mnp}$ components altogether. Interestingly if we make the following choice for $\alpha_i$:
\bg\label{sandman2}
\alpha_i~  \equiv ~ \left(1, -1, 0, 0, 0, .....\right), \nd
then we can even allow the $G_{mnpq}$ and the $G_{mnab}$ flux components to be time-independent, allowing us to rigorously realize the time scalings of the series \eqref{dqueen}, \eqref{dasecret1}, \eqref{dasecret2} and \eqref{dasecret}.   
Such a choice leads to time independent scalars in the type IIB side prompting $\vert \nabla V \vert = 0$, and therefore violating the lower bound in \eqref{vidya}. From our set of EOMs, we see that precisely in this limit most of $\alpha_i$ vanish, leading us to take into account {\it all} possible quantum corrections. This seems like a clear sign that there is no effective field theory description anymore, unless the series in \eqref{dasecret} are all convergent and so are their sum. Of course at this stage the choice \eqref{sandman2} appears adhoc, so we will have to analyze the other  EOMs to make any definitive statement.

There are two other set of time-independent EOMs that appear from analyzing the Einstein's equation related to the fiber torus ${\bf T}^2$ of $\Sigma_8$ and the three-dimensional spacetime components. They may be expressed in the following way \cite{nogo}\footnote{We do not want cross-terms in the metric of the form ${\bf g}_{M\mu}$ or 
${\bf g}_{3M}$, where $M$ are all the spatial coordinates,  to arise at the loop level as they would lead to either 
 B-fields or cross-terms in the metric in the type IIB side, ruining the de Sitter isometries.} :

{\footnotesize
\bg\label{megquaid}
&& \left(9h\Lambda + {R\over 2}\right) \delta_{ab} + {1\over 12h} \left(G_{amnp}G_{b}^{~~mnp} - {1\over 2} \delta_{ab} 
G_{cmnp}G^{cmnp}\right) + h^{1/3} \sum_{\{\alpha_i\} = 0} \mathbb{C}^{(i)}_{ab} = 0\nonumber\\
&&\left(3\Lambda + {R\over 2h} - {\square h \over 2h^2}\right) = {G_{mnpa} G^{mnpa} \over 24 h^2} + 
{\kappa^2 T_2 n_3 \over h^2 \sqrt{{\rm det}~g_{mn}}} ~\delta^8 (x - X)  - {1\over 3 h^{2/3}} \sum_{\{\alpha_i\} = 0} 
\mathbb{C}^{\mu(i)}_\mu,  \nd}
where $n_3$ is the number of static M2-branes located at any point $X \equiv (y, z, \bar{z})$ in the internal space 
$\Sigma_8$. The quantum pieces are again summed over all $\alpha_i$ that vanish, similar to what we had earlier. Note that $\mathbb{C}^{\mu(i)}_\mu$ and $\mathbb{C}_{ab}^{(i)}$ are determined by putting non-trivial metrics for 
$2+1$ dimensional spacetime and the fiber torus ${\bf T}^2$ respectively and then computing their effects on 
\eqref{gaiman}. After which one can extract $T_{\mu\nu}$ and $T_{ab}$ using \eqref{20list18}. 

The two remaining time-dependent EOMs appearing from Einstein's equations are easy to find. They are again constructed using $G_{mnpq}$ and $G_{mnab}$ fluxes and, now ignoring the $G_{0mnp}$ flux components, may be expressed in the following way \cite{nogo}:

{\footnotesize
\bg\label{churibidya}
&& {\eta_{\mu\nu}\over 24 h^2}\left({G_{mnpq}G^{mnpq}\over 4} + {G_{mnab}G^{mnab}\over \Lambda^2 \vert t\vert^4}
\right) - {1\over h^{2/3}} \sum_{\{\alpha_i\} \ne 0} \left(\Lambda \vert t \vert^2\right)^{\alpha_i - 1} \mathbb{C}^{(i)}_{\mu\nu} = 0 \\ 
&&{1\over 4h} \left(G_{acmn}G_{b}^{~~cmn} - {1\over 4}\delta_{ab} G_{mncd}G^{mncd}\right) 
- {\delta_{ab}\Lambda^2 \vert t \vert^4 \over 96 h} G^2_{mnpq} + h^{1/3} \sum_{\{\alpha_i\} \ne 0} 
\left(\Lambda \vert t\vert^2\right)^{\alpha_i + 1} \mathbb{C}_{ab}^{(i)} = 0. \nonumber \nd}
Similar issues encountered earlier for \eqref{rampling} arise again, and may be resolved by taking slowly varying flux components as before. Interestingly, we see that if we impose \eqref{sandman2}, the $G_{mnpq}$ and 
$G_{mnab}$ flux components become time independent as before.  It is somewhat miraculous that the choice 
\eqref{sandman2} allows us to choose time independent fluxes and yet get a time dependent metric of the form 
\eqref{li2018} in type IIB theory. The caveat however is the choice \eqref{sandman2} itself: allowing most of $\alpha_i$ to vanish entails {\it all} quantum corrections and, unless we have some hierarchy between the various quantum pieces, there seems to be no simple EFT description in four-dimensions.  

What about more generic choice than the one considered in \eqref{sandman2}? This is a pertinent question to ask at this stage because the choice \eqref{sandman2} is not quite motivated from the physical criteria of the theory. Let us then choose the following values for $\alpha_i$:
\bg\label{monghtpt}
\alpha_i ~ \equiv ~ \left(1, -1, a_1, a_2, a_3, ...., a_n, 0, 0, 0, ....\right), \nd
where $a_i$ are integers and we will impose some hierarchy between them. The question is how do the set of time-dependent equations \eqref{rampling} and \eqref{churibidya} behave with the choice \eqref{monghtpt}? To analyze this, note that the three set of equations in \eqref{rampling} and \eqref{churibidya} may now be written as:
\bg\label{mjackson}
\mathbb{F}^{(i)}_1(y) + \mathbb{G}^{(i)}_1(t) \mathbb{F}^{(i)}_2(y) + 
\sum_{k = 3}^{n+2} \mathbb{G}^{(i)}_{2k}(t) \mathbb{F}^{(i)}_{3k}(y) ~ = ~ 0, \nd
where $i = 1, 2, 3$ and the sum over $k$ takes into account all the $a_i$ variables appearing in \eqref{monghtpt}. By construction \eqref{mjackson} only involves all the non-zero values of \eqref{monghtpt}. The 
$\mathbb{F}^{(i)}_1$ functions are defined in the following way:
\bg\label{F11} 
&& \mathbb{F}_{1mn}^{(1)} = G_{mpqr} G_{n}^{~~pqr} - {1\over 8} g_{mn} G_{pqrl}G^{pqrl} + 12 h^{4/3} \mathbb{C}^{(1)}_{mn} \\
&& \mathbb{F}_{1\mu\nu}^{(2)} = \eta_{\mu\nu} G_{mnpq} G^{mnpq} - 24 h^{4/3} \mathbb{C}^{(1)}_{\mu\nu}, ~~~
\mathbb{F}_{1ab}^{(3)} = \delta_{ab} G_{mnpq} G^{mnpq} - 96 h^{4/3} \mathbb{C}^{(1)}_{ab}, \nonumber \nd
where the $\mathbb{C}_{MN}^{(1)}$ are the quantum pieces, for example  \eqref{dasecret}, appearing in the time dependent equations 
\eqref{rampling} and \eqref{churibidya} and the flux components are the unwarped flux components of \cite{nogo} as mentioned earlier. In the same vein, the other $\mathbb{F}_2^{(i)}$ functions take the following form:
\bg\label{julpacha}
&&\mathbb{F}_{2\mu\nu}^{(2)} = \eta_{\mu\nu} G_{mnab} G^{mnab} - 24 h^{4/3} \mathbb{C}^{(2)}_{\mu\nu} \nonumber\\
&&\mathbb{F}_{2mn}^{(1)} = G_{mpab} G_{n}^{~~pab} - {1\over 4} g_{mn} G_{pqab}G^{pqab} + 4 h^{4/3} \mathbb{C}_{mn}^{(2)}\nonumber\\
&&\mathbb{F}_{2ab}^{(3)} = G_{acmn} G_{b}^{~~cmn} - {1\over 4} \delta_{ab} G_{mncd}G^{mncd} + 4 h^{4/3} \mathbb{C}_{ab}^{(2)}, \nd
where they follow the same pattern as in \eqref{mjackson} and we have used the tensorial notations to distinguish the various functions. Finally the $F_{3k}^{(i)}$ functions take the following form:
\bg\label{poladom}
F_{3kmn}^{(1)} = 12 h^{4/3}\mathbb{C}_{mn}^{(k)}, ~~F_{3k\mu\nu}^{(2)} = - 24 h^{4/3}\mathbb{C}_{\mu\nu}^{(k)}, ~~
F_{3kab}^{(3)} = - 96 h^{4/3}\mathbb{C}_{ab}^{(k)}. \nd
All the $\mathbb{F}_m^{(i)}$ are functions of the internal coordinates of $\Sigma_8$, although for simplicity we take them to be functions of the six-dimensional base $y$ (and for some components, functions of $x_{11}$). They are also rank 2 tensors. The
$\mathbb{G}_1^{(i)}$ and $\mathbb{G}_{2k}^{(i)}$, on the other hand, are scalar functions of $t$ only and we can define them in the following way:
\bg\label{tagrathai} 
\mathbb{G}_1^{(1)} = {3\over \Lambda^2\vert t \vert^4} = 3 \mathbb{G}_1^{(2)} =  -{3\over 28} \mathbb{G}_1^{(3)}, ~~~~ \mathbb{G}_{2k}^{(1)} = \left(\Lambda \vert t \vert^2 \right)^{\alpha_k - 1} = \mathbb{G}_{2k}^{(2)} = 
\mathbb{G}_{2k}^{(3)}. \nd
Clearly the above set of functions suggests that \eqref{mjackson} is rather hard to solve analytically, so one needs to go order by order in the choice of $a_i$ in \eqref{monghtpt}. Let us first take the simplest case with $a_i = 0$. This is of course the choice \eqref{sandman2} encountered earlier.  Since $\mathbb{G}_1^{(i)}$ are not constants, the simplest solution of the system of equations are:
\bg\label{meyekechi}
\mathbb{F}_1^{(i)}  ~ = ~ \mathbb{F}_2^{(i)} ~ = ~ 0, \nd
implying that all the six equations in \eqref{F11} and \eqref{julpacha} vanish. The quantum terms $\mathbb{C}_{MN}^{(i)}$ are essential to fix the flux components, and after the dust settles, one may show that 
$G_{mnpq}$ as well as $G_{mnab}$ are time independent functions. 

Question now is whether such time independency can be maintained if we switch on $a_1$ in \eqref{monghtpt}. In the language of \eqref{mjackson} this means we are switching on $\mathbb{G}_{23}^{(i)}$ and $\mathbb{F}_{33}^{(i)}$. The class of solution for the set of equations is not hard to find, and may be expressed as:
\bg\label{pippro} 
 \mathbb{G}_1^{(i)} = a^{(i)} \mathbb{G}_{23}^{(i)} + b^{(i)}, ~~~
\mathbb{F}_1^{(i)} = b^{(i)} \mathbb{F}_2^{(i)}, ~~~ \mathbb{F}_{33}^{(i)} = -a^{(i)} \mathbb{F}_2^{(i)}, \nd
where $a^{(i)}$ and $b^{(i)}$ are constants (the repeated indices are not summed over).  Looking at the time dependences of $\mathbb{G}_1^{(i)}$ and $\mathbb{G}_{23}^{(i)}$ it is easy to infer that:
\bg\label{mcawed}
b^{(i)} = 0, ~~~a^{(1)} = 3, ~~~a^{(2)} = 1, ~~~a^{(3)} = -28, ~~~ \alpha_3 =  a_1 = -1, \nd
where we see that the extra quantum bit is fixed to $a_1 = -1$ making it coincide with the second alphabet in 
\eqref{monghtpt}. Any other choice is not allowed by the first equation in \eqref{pippro} as we want to keep 
$\Lambda$ constant. This means the $G_{mnpq}$ and $G_{mnab}$ flux components can still be time-independent and satisfy the following set of equations:
\bg\label{elistagra}
\mathbb{F}_1^{(i)} = 0, ~~~~~  \mathbb{F}_{33}^{(i)} = - a^{(i)} \mathbb{F}_2^{(i)}, \nd
where the functional forms for $\mathbb{F}_1^{(i)},  \mathbb{F}_{33}^{(i)}$ and $\mathbb{F}_2^{(i)}$ may be extracted from \eqref{F11}, \eqref{poladom} and \eqref{julpacha} respectively, and $a^{(i)}$ values are taken from 
\eqref{mcawed}. 

Switching on the other $a_i$ components in \eqref{monghtpt}, we can easily see that the analysis follows similar pattern. The $G_{mnpq}$ and $G_{mnab}$ flux components can still remain time-independent and the generic choice for 
$\alpha_i$ appears to be:
\bg\label{iiser}
\alpha_i ~ \equiv ~ \left(1, -1, -1, -1, -1, -1, ....., -1, 0, 0, 0, 0, .....\right). \nd
The ($-1$) chain suggests that the set of quantum pieces are treated equally and therefore a simple redefinition of these terms implies an equivalence to the original choice of \eqref{sandman2}, at least in the set-up that we concentrate here\footnote{The issue of sum over $\mathbb{C}_{MN}^{(i)}$ should appear here too, but since the ($-1$) chain is finite this is not as acute as having an infinite chain of ($-1$). Such a case will be discussed soon.}. This hopefully provides one additional justification for the choice \eqref{sandman2}. 

A question however arises regarding the quantum series in the time-dependent equations of \eqref{rampling} and 
\eqref{churibidya}: what if we combine the ($-1$) chain of \eqref{iiser} to express the $\alpha_i$ values, not as 
\eqref{sandman2} but as \eqref{monghtpt}? In other words  can we view the $a_i$ pieces in \eqref{monghtpt} 
to allow for the following arrangements of the quantum pieces: 
\bg\label{japtagra}
\sum_{i = 1}^n \left(\Lambda\vert t \vert^2\right)^{a_i} \mathbb{C}^{(i)}_{mn} ~ = ~ \sum_{i = 1}^N \left(\Lambda\vert t \vert^2\right)^{a_i - 1} \mathbb{C}^{(i)}_{\mu\nu} ~ = ~ 
\sum_{i = 1}^N \left(\Lambda\vert t \vert^2\right)^{a_i + 1} \mathbb{C}^{(i)}_{ab} ~ = ~ 0, \nd
so that higher order quantum effects do not change the equations of motion? Clearly such an arrangement is an attractive explanation for the choices of $\alpha_i$ in \eqref{monghtpt} instead of \eqref{sandman2} or its equivalent form \eqref{iiser}. However we now face a rather severe issue:  \eqref{japtagra} cannot quite be a reasonable explanation because each of the equations in \eqref{japtagra} has to be valid at {\it any} instant of time $t$, giving rise to a continuous infinite number of constraints. Such a constrained system doesn't appear to have any non-trivial solutions. Thus the simplicity of \eqref{iiser} or \eqref{sandman2} cannot quite be attributed to quantum cancellations of the form \eqref{japtagra}, instead we should view the whole tower of ($0$) chain to enter the time-independent equations \eqref{megrya} and \eqref{megquaid} to allow for time-independent internal metric $g_{mn}$ and internal G-flux $G_{mnpa}$. Unfortunately the quantum pieces entering \eqref{megrya} and \eqref{megquaid} do not have any apparent hierarchy\footnote{They are suppressed by $M_{p}$ but as we saw in \eqref{gaiman} and \eqref{dasecret} it is not clear whether this  supplies sufficient hierarchy between the quantum pieces.} so doesn't seem to have any simple effective field theory description from which we can extract the values of 
$g_{mn}$ and $G_{mnpa}$ components. 

The above set of constraints in \eqref{japtagra} already raises formidable problems, but we will press on by assuming that the ($a_i$) chain in \eqref{monghtpt} is somehow cancelled out. This way we only retain a chain of the form 
\eqref{sandman2}, and following this logic, 
another related question can also be asked at this stage. Instead of a semi-infinite sequence of ($0$)'s in 
\eqref{iiser} or \eqref{monghtpt}, or their equivalent form \eqref{sandman2}, can we allow for a sequence of mostly 
($-1$) in say \eqref{iiser}? For example what would happen if we allow for the following sequence for $\alpha_i$:
\bg\label{no23}
\alpha_i \equiv ~ \left(1, -1, -1, -1, -1, -1, ........, -1\right), \nd
with no zeroes appearing anywhere? One issue immediately arises with the choice \eqref{no23}: the time-independent equations for the metric and the fluxes, i.e  \eqref{megrya} and \eqref{megquaid} respectively, do not have any contributions from the quantum pieces. Such a system of equations 
cannot have any solutions, as we shall argue in section \ref{nala}. This clearly rules out the choice \eqref{no23}.

However let us assume, just for the sake of an argument, that a choice like \eqref{no23} somehow manages to provide consistent solutions for the system of equations in \eqref{megrya} and \eqref{megquaid}. Question then is: will 
\eqref{no23} allow for an effective field theory description in lower dimensions? Plugging \eqref{no23} in the set of equations \eqref{rampling} and \eqref{churibidya}, we see that the quantum pieces in these equations make the following contributions:
\bg\label{vmadsen}
{\sqrt{h}\over g_s^2}\left(\sum_{i = 2}^{\infty} \mathbb{C}_{MN}^{(i)}\right), \nd
where $h(y)$ is the warp-factor and $g_s$ is the type IIA coupling \eqref{cashjul}. The sum of the quantum pieces are arranged without any hierarchies and it is not clear, for example from \eqref{dasecret}, whether the sum of the 
$\mathbb{C}_{MN}^{(i)}$ terms is convergent. A similar question could also arise for the quantum sums in \eqref{megrya} and \eqref{megquaid}: the convergence of each of the series is not clear at all. Additionally, and it is one of the most pertinent observation, all the quantum series of $\mathbb{C}_{MN}^{(i)}$ appear in the sum \eqref{vmadsen} with {\it equal} footings. This is of course the root of our problem regarding the absence of an effective field theory description of the system. One possible way out is to Borel sum the series in \eqref{megrya} and \eqref{megquaid} (and also in \eqref{vmadsen}) to make sense of them.  If doable, Borel summing will imply introducing infinite degrees of freedom ruining a simple effective field theory description of the theory. Whether such a procedure is indeed possible has never been checked. Thus more work is needed to make sense of the quantum effects in this theory.

Let us now come to a more generic analysis of the quantum series. The issue raised with the set of constraints 
in \eqref{japtagra} could in fact be related to our choice \eqref{oshotmeye}, which is simply not generic enough to accommodate such large set of constraints.
One way out would be to maybe 
identify $\mathbb{C}_{MN}^{(k)}$ in 
\eqref{20list18} to some linear combinations of $\mathbb{C}_{MN}^{(\eta, \gamma)}$ in 
\eqref{dasecret3}. Implementing this will then lead to a new possibility of {\it rearranging} the quantum series for $T_{MN}$ in a different way from what we did in \eqref{rakth}. How should this be done?
The time neutral series  \eqref{dasecret} gives us a hint as to how to proceed. Using these,  let us define the rank 2 tensor $\mathbb{C}^{(k)}_{MN}$ of \eqref{20list18} in the following way:
\bg\label{mariaC}
\mathbb{C}^{(k)}_{MN} ~\equiv ~ \sum_{l} \alpha^{(k)}_{l} \mathbb{H}^{(l)}_{MN}, ~~~~ {\rm where} ~~
\mathbb{H}^{(l)}_{MN} \equiv M_p^2 \mathbb{C}_{MN}^{(\eta, \gamma)}, ~~ l \equiv \vert a_\eta\vert - 
\vert a_\gamma\vert, 
\nd  
and $\alpha_{l}^{(k)}$ are dimensionless numbers or more generically dimensionless Lorentz invariant functions of the unwarped curvature and fluxes. The other function appearing above, namely
$\mathbb{C}_{mn}^{(\eta, \gamma)}$ is given in \eqref{dasecret3} with similar constructions  for 
$\mathbb{C}_{ab}^{(\eta, \gamma)}$ and $\mathbb{C}_{\mu\nu}^{(\eta, \gamma)}$. We have assumed that 
$\alpha_{l}^{(k)} \ne \alpha_{l}^{(m)}$, and therefore different values of $k$ in 
\eqref{mariaC} will represent different combinations of the time neutral series 
in say \eqref{dasecret3}. We can now use these time-neutral $\mathbb{C}_{MN}^{(k)}$ functions to define an energy momentum tensor of the form:
\bg\label{hochelaga}
 T_{MN} \equiv \sum_k \left(\Lambda\vert t\vert^2\right)^{\beta_k} \mathbb{C}_{MN}^{(k)}, \nd
 which is a variant of the energy-momentum tensor given earlier in \eqref{rakth}. The story now proceeds in exactly the same way we studied before. For example, the choice \eqref{sandman2} would now tell us that the chain of ($0$)'s are precisely the sum of the $\mathbb{C}_{MN}^{(k)}$ from \eqref{mariaC}, which in turn are the various combinations of the time neutral series 
 $\mathbb{H}_{MN}^{(l)}$ being summed over. Such a construction is clearly more involved than the simple picture that we had before, but shares the same flavor of problems that we encountered earlier, albeit now in a different guise. For example, previously we had summed over all the $\mathbb{C}_{mn}^{(\eta, \gamma)}$ functions to analyze the time independent EOM \eqref{megrya}. The $M_p$ hierarchy in each $\mathbb{C}_{mn}^{(\eta, \gamma)}$ in the end is clearly a red herring and is therefore irrelevant to our discussion here. The fact that {\it all} $\mathbb{C}_{mn}^{(\eta, \gamma)}$ appeared equally was the root of the problem.  In the present case, with a redefined quantum sum, the issue is more acute: all 
 $\mathbb{C}_{mn}^{(k)}$ would now appear equally creating the same issue as before, albeit more strongly. It is then the {\it sum} of the ({sum}-of-the) $\mathbb{H}_{mn}^{(l)}$ functions that form the root cause of problems.   
 
 At this point we could also entertain a chain of the form \eqref{monghtpt}, and in turn ask similar questions as before. Much like \eqref{japtagra} the situation at hand will give rise to similar set of constraints
again leading to continuous infinite number of constraints, possibly forbidding a non-trivial solution. The story then seems almost similar to what we had before with two exceptions. The first, can be seen from the fact that  previously the $\mathbb{C}_{MN}^{(k)}$ from 
\eqref{dasecret}
were all different functions of ${\bf R}$ and ${\bf G}$, but now the $\mathbb{C}_{MN}^{(k)}$ from \eqref{mariaC} could become similar if 
$\alpha_l^{(k)} = \alpha_l^{(m)}$ for $k \ne m$. The second, appears from a choice of a set of $\alpha^{(k)}_l$ that lead to vanishing $\mathbb{C}_{MN}^{(k)}$ with non-vanishing $\mathbb{H}_{MN}^{(l)}$ ingredients. These are precisely the set of 
$\mathbb{C}_{MN}^{(k)}$ that may be used to satisfy the set of constraints in \eqref{japtagra} via:
\bg\label{renaissance}
\mathbb{C}_{mn}^{(p)} ~ = ~ \mathbb{C}_{ab}^{(p)} ~ = ~ \mathbb{C}_{\mu\nu}^{(p)} ~ = ~ 0, \nd
for certain set of $k = \{p \}$, i.e for certain linear combinations of $\mathbb{H}_{MN}^{(l)}$ with the set of 
$\{\alpha^{(p)}_l\}$.
This way we see that \eqref{renaissance} may now provide non-trivial solutions to the equations of motions without violating the 
no-go conditions of \cite{nogo}, provided we retain a set of  $\alpha_l^{(k)}$  {\it different} from 
$\alpha_l^{(m)}$ when $k \ne m$ giving the required non-vanishing $\mathbb{C}_{MN}^{(k)}$. These 
$\mathbb{C}_{MN}^{(k)}$ are the  time-neutral rank 2 tensors that are all different functions of ${\bf R}$ and 
${\bf G}$ appearing for example in the EOMs \eqref{megrya} and \eqref{megquaid}. The detailed structure is elucidated in {\bf Table \ref{japnunur}}. We can now see how the original choice of \eqref{oshotmeye} fits in the 
 generalized picture. Consider the following choice of the coefficients $\alpha_l^{(\pm k)}$:
 \bg\label{dhakka}
 \alpha^{(\pm k)}_l ~ = ~ \delta^{(\pm k)}_l, ~~~~~ k = 1, ......, n, \nd
 which can be plugged in the EOMs \eqref{megrya}, \eqref{megquaid}, \eqref{rampling} and \eqref{churibidya} leading to the constructions that we had earlier. The difference however arises once we start analyzing the set of constraint equations \eqref{japtagra}. 
 The above identification \eqref{dhakka} is not helpful for $k > n$. In fact for $k > n$, we should resort back to the linear combinations \eqref{mariaC} to allow for a consistent solution. Notice that in {\bf Table \ref{japnunur}} we have put an 
 upper limit of $\vert k\vert = \vert s\vert$. This is because both $n$ and $s$ are arbitrarily large as the number of 
 time-neutral rank 2 tensors in \eqref{dasecret3} are arbitrarily large (with each having further infinite degeneracies). 
 Thus from the generalized construction \eqref{mariaC}, many different arrangements of the quantum terms may be made for $k \le n$. 
 
In the end however, as we discussed earlier, none of the above constructions can save the day because of the underlying loss of $g_s$ hierarchy. Therefore a different rearrangements of the quantum terms cannot quite help us in resolving the root cause of the problem, although it does help us in giving a consistent class of solutions with the constraints \eqref{japtagra}. 

\begin{table}[h!]
 \begin{center}
\begin{tabular}{|c|c|c|}\hline Coefficients $\forall ~l$  & Time neutral rank 2 tensors & Relevant equations  \\ \hline
$\alpha_l^{(\pm 1)}$ & $\mathbb{C}^{(\pm 1)}_{MN}$  & \eqref{rampling}, \eqref{churibidya}   \\  \hline
$\alpha_l^{(0)}, \alpha^{(\pm 2)}_l, ...., \alpha^{(\pm n)}_l$ & $\mathbb{C}^{(0)}_{MN}, \mathbb{C}^{(\pm 2)}_{MN}, ...., \mathbb{C}^{(\pm n)}_{MN}; ~~ 
\mathbb{C}_{MN}^{(0)} + \sum^n_{k = 2} \mathbb{C}^{(\pm k)}_{MN}$  & \eqref{megrya}, 
\eqref{megquaid} \\  \hline
$\alpha^{(\pm n \pm 1)}_l, ...., \alpha^{(\pm n \pm s)}_l$ & $\mathbb{C}^{(\pm n \pm 1)}_{MN}  = \mathbb{C}^{(\pm n \pm 2)}_{MN} = ... = \mathbb{C}^{(\pm n \pm s)}_{MN} = 0$   
& \eqref{japtagra}  \\  \hline
  \end{tabular}
\end{center}
  \caption{The full contributions from the quantum effects in the energy-momentum tensor to the EOMs. Since both 
  $n$ and $s$ can be arbitrarily large, the second row elucidates the loss of an effective field theory description, whereas the third row provides an exact solution to the constraint equations.}
\label{japnunur}
\end{table}


Once we make the $G_{mnpq}$ and $G_{mnab}$ fiux components time dependent, in turn providing an alternative resolution of the tension we had in realizing \eqref{monghtpt} and \eqref{japtagra}, the situation at hand changes quite a bit. For example let us consider the following behavior of the flux components:
\bg\label{888}
&&G_{mnpq}(t, y) ~\equiv ~ \sum_i \left(\Lambda\vert t\vert^2\right)^{-\vert a_i\vert} g^{(i)}_{mnpq}(y) \nonumber\\
&&G_{mnab}(t, y) ~\equiv ~ \sum_i \left(\Lambda\vert t\vert^2\right)^{-\vert b_i\vert} g^{(i)}_{mnab}(y), \nd
where the signs of the exponents are chosen such that the fluxes decrease as  we approach early times\footnote{More generic choices are clearly possible, and we will discuss implications of them later.}. 
We have also assumed the fluxes to be defined on the six-dimensional base parametrized by $y$, and 
$g^{(i)}_{mnpq}$  and $g^{(i)}_{mnab}$ are some functions of $y$ to be determined from the two equations \eqref{rampling} and \eqref{churibidya} as well as the G-flux equation:
\bg\label{rumpi}
&&D_M\left({\bf G}^{MNPQ}\right) =  {1\over \sqrt{-{\bf g}}} \epsilon^{NPQM_1......M_8}\left[{1\over 2(4!)^2} 
{\bf G}_{M_1...M_4} {\bf G}_{M_5...M_8} + {2\kappa^2 T_2\over 8!} \left({\bf X}_8\right)_{M_1....M_8}\right] \nonumber\\
&&~~~~~~ + {2\kappa^2 T_2n_3\over \sqrt{-{\bf g}}} \int d^3\sigma~\epsilon^{\mu\nu\rho} \partial_\mu X^N 
\partial_\nu X^P \partial_\rho X^Q ~\delta^{11}(x - X) + {1\over \sqrt{-{\bf g}}} \left({\delta S_{quantum}\over \delta 
{\bf C}_{NPQ}}\right), \nd
where 
$\bf{G}_{MNPQ}$  and $\bf{C}_{MNP}$ are the {\it warped} components in the sense that the indices are raised or lower by the warped M-theory metric (and as such involve time-dependent  pieces).
Similarly the determinant of the metric ${\bf g}$ is the warped M-theory metric and it is used to define the covariant derivative 
$D_M$. However the epsilon tensor is defined using the 
{\it unwarped} metric components and $T_2$ gives the tension of the M2-branes.
As such $\kappa^2T_2$  takes care of the dimensions of the topological and the brane terms (see \cite{nogo} for more details). 

Before moving ahead with the analysis of \eqref{888}, let us verify the consistency of our original choice of 
\eqref{cold} which was derived from the slow moving membranes. Since the membranes are still expected to move slowly, the choice \eqref{cold} should remain a valid choice now too. Plugging \eqref{cold} in \eqref{rumpi} then gives us:
\bg\label{parkbedard}
-\square h & = & {1\over 12} G_{mnpa} \left(\ast_8 G\right)^{mnpa} + {2\kappa^2 T_2 \over 8! \sqrt{g}} \left(X_8\right)_{M_1.....M_8} \epsilon^{M_1.....M_8} \\
&& ~~~+ {2\kappa^2T_2 \over \sqrt{g}} \Big[ n_3 \delta^8(x - X) - \bar{n}_3 \delta^8(x - Y)\Big] + 
{1\over \sqrt{g}}\left( {\delta S_{top}  \over \delta {\bf C}_{012}} + {\delta  S_{ntop} \over \delta {\bf C}_{012}}\right), \nonumber
\nd
where the result is expressed in terms of unwarped metric and flux components with ($n_3, \bar{n}_3$) being the number of M2 branes and anti-branes respectively where, for simplicity, we can assume that they differ by 1 to not change the LHS of \eqref{parkbedard}.  We have also divided the quantum corrections to topological and the non-topological pieces, same way as we did in \cite{nogo}. 

There are two interesting points to note from \eqref{parkbedard}. First, the flux components $G_{mnpa}$ continues to be time independent even if $G_{012m}, G_{mnpq}$ and $G_{mnab}$ are all time dependent as in 
\eqref{cold} and \eqref{888} respectively. Secondly, the warp factor $h$ will be a smooth function in the fully quantum corrected scenario (more on this in the next section), and therefore integrating the LHS of \eqref{parkbedard} over the eight manifold $\Sigma_8$, assuming no boundary, is expected to reproduce the anomaly cancellation condition with M2-branes and fluxes of \cite{DM, DRS}. Subtlety however arises because the metric of $\Sigma_8$ is time dependent, whereas the analysis of \cite{DM, DRS} are exclusively for static backgrounds. The ${\bf X}_8$ polynomial is topological, so doesn't quite depend on how the metric of ${\bf X}_8$ changes with time, but the quantum corrections are heavily constrained because:
\bg\label{alice}
{\partial\over \partial t}\left( {\delta S_{top}  \over \delta {\bf C}_{012}} + {\delta  S_{ntop} \over \delta {\bf C}_{012}}\right) = 0. \nd
Whether this is possible to maintain remains to be seen. One simple solution could be to take the terms in the bracket of \eqref{alice} vanishing, in which case the anomaly cancellation condition of \cite{DM, DRS} will not change. 
Another solution could be that the contributions to the bracket come from Lorentz-invariant time-neutral pieces, similar to the ones in \eqref{dqueen}, but now with spacetime flux components $G_{012m}$.  
Such contributions are difficult to construct in practice , so it remains to be seen how a scenario like this might be realized in our set-up. Clearly more work is needed to make any definitive statement here. 

Coming back to the G-flux components \eqref{888}, it is now easy to see that the $\alpha_i$ values can be different from what we had in \eqref{sandman2} or \eqref{iiser}. We can take the following chain of alphabets for $\alpha_i$:
\bg\label{cornbaby}
\alpha_i ~\equiv ~ \left(a_1, a_2, a_3, ....., a_n, 0, 0, ....., 0, 0, a_{n+1}, a_{n+2}, ......\right), \nd
where the finite chain of (0) is there to maintain the time independent flux components $G_{mnpa}$. This is necessary otherwise the anomaly cancellation condition will get even more constrained from the integral of 
\eqref{parkbedard} over $\Sigma_8$. In the language of {\bf Table \ref{japnunur}} it is as though 
the first row has expanded to accommodate the ($a_i$) chain from \eqref{cornbaby}, and the  second row has substantially reduced. That this is possible, despite the fact that we have an infinite number of time-neutral rank 2 tensors, may be seen from the fact that any $\alpha_l^{(\pm k)}$ that goes in the first row of {\bf Table \ref{japnunur}}
cannot reappear in the time-neutral series in the second row to avoid double-counting under the choice
\eqref{dhakka}. This means we can use most of the $\alpha_l^{(\pm k)}$ coefficients to solve the time-dependent equations \eqref{rampling} and \eqref{churibidya}, leaving a finite chain of time-neutral pieces behind for the time-independent equations \eqref{megrya} and \eqref{megquaid}. These finite chain of ($0$) tell us that the sum of the quantum pieces: 
\bg\label{duris} 
\sum_{\{\alpha_i\} = 0} \Big\{\mathbb{C}_{mn}^{(i)}, \mathbb{C}_{ab}^{(i)}, \mathbb{C}_{\mu\nu}^{(i)}\Big\}, \nd 
can be controlled provided the individual pieces 
$\mathbb{C}_{MN}^{(i)}$ themselves have convergent series. On the other hand, the quantum series in 
\eqref{rampling}  and \eqref{churibidya} tell us that the $g_{mnpq}^{(i)}$ and $g_{mnab}^{(i)}$ components from 
\eqref{888} will now be determined in terms of the semi-infinite sum of quantum series $\mathbb{C}_{MN}^{(i)}$ provided the series have a well defined hierarchy. 
Therefore the whole analysis now revolves not just around around the convergence of the series in 
\eqref{mariaC}, but more importantly on the existence of a hierarchy. Whether this is possible or not will be the subject of the following section.

\subsection{Quantum constraints, hierarchy and the swampland \label{nala}} 

The quantum constraints that we discussed in the previous section, both for the time independent as well as the time dependent cases, can now be succinctly presented by combining the three equations in \eqref{megrya} and 
\eqref{megquaid} in the following way \cite{nogo}:
\bg\label{relcha}
&& {1\over 12} \int d^8 x \sqrt{g}~ G_{mnpa}G^{mnpa} + 12 \Lambda \int d^8 x \sqrt{g}~h^2 + 2\kappa^2 T_2 n_3 
\nonumber\\
&& ~~~ + \int d^8 x \sqrt{g} h^{4/3}\left({1\over 2} \sum_{\{\alpha_i\} = 0} \mathbb{C}^{a, i}_a + 
{1\over 4} \sum_{\{\alpha_i\} = 0} \mathbb{C}^{m, i}_m - {2\over 3} \sum_{\{\alpha_i\} = 0} \mathbb{C}^{\mu, i}_\mu\right) ~ = ~ 0, \nd
where $\Lambda > 0$ and only the time-independent flux components $G_{mnpa}$ appear in the above equation, even if we have other time-independent flux components. The second line is the contribution from the quantum pieces 
that we discussed earlier, and the above equation should be regarded as a {\it constraint} on the quantum pieces because all the terms in the first line are positive definite. Therefore the above equation will only have a solution if the following constraint is satisfied \cite{nogo}:
\bg\label{chitchor}
   {1\over 2} \sum_{\{\alpha_i\} = 0} \langle \mathbb{C}^{a, i}_a\rangle + 
{1\over 4} \sum_{\{\alpha_i\} = 0} \langle \mathbb{C}^{m, i}_m \rangle 
- {2\over 3} \sum_{\{\alpha_i\} = 0} \langle \mathbb{C}^{\mu, i}_\mu \rangle ~ < ~ 0, \nd  
where the expectation values are defined by simply integrating the $h^{4/3}$ weighted quantum pieces over the eight-manifold as in \cite{nogo}. The above constraint is highly non-trivial\footnote{The fact that this constraint has 
{\it no} solution in the presence of branes, anti-branes, orbifold and orientifold planes as well as the $p$-form fluxes in the {\it absence} of the quantum corrections has already been discussed in \cite{nogo} so we will not elaborate the story anymore. Interested readers may find all the details in \cite{nogo} (and verify for himself or herself that for the choice of $\alpha_i$ in \eqref{no23} no solutions exist). Instead we want to concentrate on the convergence and the hierarcy issues of the quantum series here.}
not only because it involves the 
$\mathbb{C}_{MN}^{(i)}$ factors from \eqref{mariaC}, where the thorny issue of convergence would reappear, but also because now it involves a {\it sum} of all the $\mathbb{C}_{MN}^{(i)}$ factors arranged so that it is a negative definite number. Could such constraint be ever satisfied? 

To answer all the questions raised above, let us study the scenario at hand more carefully. First note that we are in principle talking of {\it four} different series here. They can be tabulated in the following way:

\vskip.1in

\noindent (1) The series  $V_{\mathbb{Q}}$ given in \eqref{gaiman} which is the main series of  quantum corrections, and is expressed with inverse powers of $M_p$. 

\vskip.1in

\noindent (2) The series of time-neutral functions, some of which are presented in \eqref{dqueen}. They are also expressed as inverse powers of $M_p$.

\vskip.1in

\noindent (3) The complete series of $\mathbb{C}_{MN}^{(i)}$ functions, given in \eqref{mariaC}, and expressed in inverse powers of $M_p$. This series involve the time-neutral series as a subset, plus they are infinitely degenerate.

\vskip.1in

\noindent (4) The sum of all the series $\mathbb{C}_{MN}^{(i)}$ which are collected for example from the chain of ($0$)'s in \eqref{sandman2}.  These ($0$)'s are arranged in a semi-infinite chain. 

\vskip.1in

\noindent The convergence of each of these series is important to make sense of all the quantum corrections in the theory. Since all the pieces in every series are suppressed by inverse powers of $M_p$,  clearly there seems to be at least one simple way to control them: take the $M_p \to \infty$ limit.  Unfortunately this simple procedure doesn't work because of various factors including  time-dependences of the M-theory metric as well as the presence of non-local and non-polynomial pieces in the quantum corrections (that we didn't discuss) which may go as positive powers of 
$M_p$.\footnote{As an example, let us consider any of the $\Lambda_{(k)}$ pieces in \eqref{dqueen}. They are all time-neutral, and for our purpose we can choose $\Lambda_{(1)}$ as a representative. Using this let us define the following function: $$M_p^6 \int_{0}^{y_1}\int_0^{y_2}....\int_0^{y_8} d^8 y' \sqrt{{\bf g}_8}~ \square \Lambda_{(1)}(y'_1, ..., y'_8) ~\equiv ~
M_p^3 \Gamma_{(5)}(y_1, ..., y_8)$$
\noindent which is by construction a time-neutral function also, but now appears with a positive power of $M_p$. We can raise this to arbitrary powers to generate positive powers of $M_p^3$. By construction they are non-local functions and may therefore contribute to the non-local counter-terms discussed for example in \cite{douglas, modesto}. \label{stepgran}}.  
This time dependence in fact triggers off the type IIA coupling $g_s$, as given in \eqref{cashjul}, and therefore the suppressions factors in each of the components of a given series are not inverse powers of $M_p$, but the following combinations:
\bg\label{aletex}
{\left(\Lambda \vert t\vert^2\right)^{\pm \vert a\vert} \over M_p^b} ~ \propto ~ {\left(g_s\right)^{\pm 2\vert a \vert}\over 
M_p^b}; ~~~~~ a \ge 0, ~b \in  \mathbb{Z}, \nd
where the time-neutral series is exactly the $a = 0$ limit of \eqref{aletex}, and we have ignored the warp-factor $h$  dependence of $g_s$ in \eqref{cashjul} as this doesn't effect the results.   What is important is the appearance of both $g_s$ and $M_p$, and therefore calls for an hierarchy between them\footnote{The negative powers of $g_s$ imply non-perturbative contributions near $g_s \to 0$. As such they could be expressed as 
${\rm exp}\left[-\left({1\over g_s^{\vert a \vert}}\right)\right]$ and the quantum series could be summed accordingly. Such a conclusion can arise from Borel summing the series in $g_s$. Once the series appears non-convergent or asymptotic, Borel summability can be applied and the final answer provides a hint as to what non-perturbative effects contribute.  However 
near strong coupling, i.e $g_s \to \infty$, the $1/g_s$ effects are perturbative, so the quantum series could involve polynomial powers of 
${1\over g_s^{\vert a \vert}}$. An example is the choice \eqref{888}.\label{granny}}. 

Let us now ask, under what conditions do we expect to see the full de Sitter isometries in the type IIB side? Clearly this would happen when all the flux components are time independent. The flux components $G_{mnpa}$ are already time-independent, as we can easily infer from \eqref{megrya} and \eqref{megquaid}, and it is not very hard to make the equations for the other two components $G_{mnpq}$ and $G_{mnab}$, namely \eqref{rampling} and \eqref{churibidya}, time independent by choosing the chain \eqref{sandman2} or \eqref{iiser}. The spacetime component of the G-flux, namely the $G_{012m}$ component in \eqref{cold} is time-dependent but this is proportional to the volume form in the type IIB side and therefore respects the de Sitter isometries. We should then ask if solutions are possible at all times. If we take a finite value for $M_p$, the issue of the convergence for the infinitely degenerate series in \eqref{dasecret} reappear. However even if convergence of each of the series in \eqref{mariaC} is guaranteed, the convergence of the sum over all the $\mathbb{C}_{MN}^{(i)}$  is not clear. More so, the acute issue of the possibility to maintain the constraint \eqref{chitchor}, where the sum of three such different series appear, is clearly not guaranteed unless of course there exists a very strong hierarchy between all the quantum pieces in 
\eqref{mariaC}.

Let us elaborate this a bit more. Our careful study in the previous section told us that the quantum series that we have to consider include all the $\mathbb{C}_{MN}^{(i)}$ pieces that may be summed in the following 
fashion:

{\footnotesize
\bg\label{tiadd}
\sum_k \left(\Lambda\vert t\vert^2\right)^{\alpha_k} \mathbb{C}_{MN}^{(k)} &=& \sum_k g_s^{2\alpha_k}
\left(\sum_b {\left(c_{kb}\right)_{MN}\over M_p^{\beta_{kb}}}\right) \\
&= &  g_s^{2\alpha_1}\left[{\left(c_{11}\right)_{MN}\over M_p^{\beta_{11}}} + {\left(c_{12}\right)_{MN}\over M_p^{\beta_{12}}} + ......\right] + g_s^{2\alpha_2}\left[{\left(c_{21}\right)_{MN}\over M_p^{\beta_{21}}} + {\left(c_{22}\right)_{MN}\over M_p^{\beta_{22}}} + ......\right] + ...... \nonumber\nd}
where $\beta_{kb} \in \mathbb{Z}$; and the $\left(c_{kb}\right)_{MN}$ pieces may be extracted from \eqref{dasecret} and from the equivalent expansions for 
$\mathbb{C}_{ab}^{(k)}$ and $\mathbb{C}_{\mu\nu}^{(k)}$. The above series shows that for any given power of $g_s$, there exists a series in all powers of $M_p$ coming from, say, \eqref{mariaC}, with the non-local contributions included in. This series has a {\it weak} hierarchy governed by the $\left(c_{kb}\right)_{MN}$ factors, which in turn are functions of ${\bf R}$ and ${\bf G}$, so cannot be arbitrarily tuned. Additionally, as we saw from \eqref{mariaC}, the series also has an infinite degeneracy, but we will ignore this for the time being. In fact, as alluded to earlier, M-theory may allow us to chose a particular combination dictated by the underlying structure of the local and the non-local quantum corrections. On the other hand, the $g_s$ provides a {\it strong} hierarchy because it can be partially tuned by changing $t$ (recall, from \eqref{mariaC}, that $\left(c_{kb}\right)_{MN}$ are all time independent functions). Thus there are at least two levels of convergences that we seek here: one, the convergence of the series in $\left(c_{kb}\right)_{MN}$ and two, the convergence of the series in $g_s$. Although none are guaranteed here, the subtlety lies elsewhere. This can be seen in the following way. First, what we actually need is not the $g_s$ expansion, but the series when
$\{\alpha_k\} = 0$, which are precisely the time-neutral series that would contribute to the quantum corrections here. This implies:
\bg\label{allislost}
\sum_{\{\alpha_k\} = 0} \mathbb{C}_{MN}^{(k)} &=& 
\sum_k \left(\sum_b {\left(c_{kb}\right)_{MN}\over M_p^{\beta_{kb}}}\right) \\
&=& \left[{\left(c_{11}\right)_{MN}\over M_p^{\beta_{11}}} + {\left(c_{12}\right)_{MN}\over M_p^{\beta_{12}}} + ......\right] + 
\left[{\left(c_{21}\right)_{MN}\over M_p^{\beta_{21}}} + {\left(c_{22}\right)_{MN}\over M_p^{\beta_{22}}} + ......\right] + .....,
\nonumber \nd
where $\beta_{kb} \in \mathbb{Z}$. Looking at the each of the series in brackets in \eqref{tiadd} and \eqref{allislost}, one might 
erroneously think that there is a leading order term in each of them. However this is not the case as is evident from the following argument.
If we only consider the polynomial corrections to the action, namely \eqref{gaiman}, then each 
$\mathbb{C}^{(k)}_{MN}$'s would have a different leading power of $1/M_p$, creating a hierarchy between the quantum corrections. However as we mentioned earlier, there are also non-local corrections with positive powers of $M_p$ (see for example footnote \ref{stepgran}), so the series don't have a leading order term, potentially destroying this hierarchy. Preserving the $M_p$ hierarchy in the presence of these corrections would require that the higher order non-local corrections are sufficiently suppressed by the coefficients $\left(c_{kb}\right)_{MN}$, which cannot be guaranteed without a more detailed analysis of these terms.

The subtlety should be clear now: there are no more hierarchies left between the individual series in the brackets above. Each of the series in the brackets, which are  basically $\mathbb{C}_{MN}^{(k)}$, come with all powers of
$M_p$ and now contribute {\it equally} to the sum! The hierarchies provided by $\left(c_{kb}\right)_{MN}$, as mentioned above, are pretty weak to allow for any controlled approximation for the sum. The situation however improves dramatically once a small time dependence is switched on. Factors of $g_s$ appear as in \eqref{tiadd}, providing strong hierarchy, and in turn controlling the sum.  

Thus our  argument here implies that there is at least no simple effective field theory description for the background with full de Sitter isometries.
Of course it is always possible that we are ignoring other terms in the quantum series that would in fact allow solutions with de Sitter isometries to exist. Such quantum pieces may not be expressible as polynomial powers of ${\bf R}$ and ${\bf G}$ at weak curvatures and weak field strength, and also probably not as the non-local terms alluded to earlier. 
We are not aware of these terms, but if they occur and indeed also allow for solutions to exist, the whole swampland criteria will have to be revisited.

What happens with time dependent fluxes of the form \eqref{888}? For such a case de sitter isometries should be visible when $t \to -\infty$. Unfortunately in this limit $g_s \to \infty$ and therefore from the scaling argument in 
\eqref{aletex} we can easily see that even in the limit of finite $M_p$ or $M_p \to \infty$, unless there is well defined hierarchy between 
$g_s$ and $M_p$, none of the above four series seems convergent!\footnote{We could instead take {\it positive} 
exponents in \eqref{888}. For this case de Sitter isometries will be visible at late times where $g_s \to 0$. In either case, it is the perturbative expansion in $g_s$ that mostly matters here.} 

What happens at later time when $t$ is finite with the time-dependent fluxes? In this limit $g_s$ can be made small and, allowing a finite $M_p$, there appears some hope of controlling the quantum series in some meaningful way, and solutions could exist. However in this limit the background doesn't have all the de Sitter isometries so we will {\it not} be violating the swampland conjecture of \cite{vafa1}. Additionally due the presence of the non-zero  time dependent flux components $G_{mnpq}$ and $G_{mnab}$, there would be time-dependent moduli in four-dimensions (in the type IIB side). As we saw in \eqref{vidya}, such time-dependent moduli allow solutions to exist without violating the swampland conjecture.   

A related question would be: what about later time, i.e when $t \to 0$? Unfortunately in this limit the G-flux components in \eqref{888}  
 blow up so our simple analysis cannot provide any definitive statement here. However as hinted earlier, more generic choices of flux configurations are possible that allow finite values at late times. For such configurations, we are exploring the $g_s \to 0$ limit and solutions could  exist provided:
\bg\label{birjoani}
{g_s^{2\vert a\vert} \over M_p^b} ~ << ~ 1, \nd
in the limit when $M_p \to \infty$ and assuming the $g^{-2\vert a\vert}_s$ effects are controlled 
non-perturbatively (see footnote \ref{granny}). 
This is not much of a surprise because, as discussed above, the background does not have all de Sitter isometries, so there could be solutions without violating the swampland criteria.      

Finally, what about flat and AdS spaces? Are there effective field theory descriptions for such cases? The simplest answer is the following. The no go constraint \eqref{chitchor}, assuming this to be more generic than being derived from a metric \eqref{li2018}, can be easily solved when $\Lambda = 0$ or 
$\Lambda < 0$ {\it without} introducing the series of quantum corrections. Therefore both flat and AdS spaces are possible {\it classically}. A more non-trivial question however is the one associated with the series of quantum corrections. Do the quantum corrections allow effective field theories associated with flat and AdS 
spaces\footnote{According to the swampland criteria, for both cases $\vert \nabla V\vert = 0$. However for the flat space, $V = 0$, whereas for the AdS space, $V < 0$. The swampland criteria are clearly satisfied for both cases, allowing effective field theories to exist.}? To answer this let us bring back the two quantum series discussed in \eqref{tiadd} and \eqref{allislost}. Note that the apparent non-existence of an effective field theory for a de Sitter space, in the type IIB side, has absolutely nothing to do with the convergence or the divergence of each of the series (in powers of $M_p$) in the brackets of \eqref{tiadd} and \eqref{allislost}. The actual reason therein was the {\it loss} of $g_s$ hierarchy: each of the brackets contributed equally to the sum, and therefore an infinite collections of the brackets were taken into account to make sense of the EOMs, thus ruining an effective field theory description.
However the situation changes drastically once we take flat and AdS spaces. The four main equations, the two time-independent  equations \eqref{megrya} and \eqref{megquaid}  and the two time-dependent equations \eqref{rampling} and \eqref{churibidya}, now no longer appear in the way they appeared here.  In other words, the decoupling of the EOMs into time-dependent and time-independent pieces, solely because of the 
$\Lambda \vert t\vert^2$ factor in the metric \eqref{li2018}, does not happen now!
Additionally, the type IIA string coupling is no longer time dependent as in
\eqref{cashjul} here. This way $g_s$ can be made arbitrarily small, thus ignoring both perturbative and non-perturbative quantum corrections altogether. In the language of our de Sitter computation, this is as though we are not resorting to the
($0$) chains in the time-independent equations \eqref{megrya} and \eqref{megquaid}, and keeping $g_s \to 0$ 
in the time-dependent equations, essentially eliminating the ($-1$) chains in \eqref{rampling} and \eqref{churibidya}
altogether.    

At this stage it might be interesting to compare the situation at hand with the strong coupling behavior of QCD. The QCD action is exact, analogous to the string world sheet action. The supergravity action we play with emerges only as a limit of string theory, somewhat analogous to chiral perturbation theory as an EFT of the pion (which is valid in the hadronic phase of QCD). In the QCD case, when quantum corrections to the pion EFT become large, it indicates that the pions split into quarks, in which case one must use the full QCD action. We refer to this phenomena as the ``breakdown of effective field theory".

In our work, the quantum corrections at play are string-theoretic (e.g. string-loop diagrams), which add terms to the supergravity effective action. By ``breakdown of effective field theory", we mean the solution is no longer well described as a supergravity limit, but is intrinsically stringy.  Our work is an attempt to make progress in understanding this phase of theory utilizing both supergravity and stringy ingredients.

Let us elaborate this in some more details as it would illustrate the difference between the two approaches. Thus in more quantitative terms, for QCD the quantum corrections have the form:  
\bg\label{naak1}
{\cal Q}_1 \equiv \sum_a (g^2_{YM})^a \mathbb{C}_a, \nd
where $C_a$ are computed from loops and for $a < 0$ we have the non-perturbative (NP) effects. They eventually may be expressed as powers of $e^{-1/g^2_{YM}}$. On the other hand, for the de Sitter case the series we expect is:
\bg\label{naak2}
{\cal Q}_2 \equiv \sum _a g_s^a \mathbb{C}_a, \nd
where $\mathbb{C}_a = \sum_{b} \mathbb{D}_{ab} M_p^{b}$ and $b$ is summed for all positive and negative integers in a way discussed in details earlier. If this is the case, EFT {\it is} defined and would match with the EFT of QCD mentioned above. Unfortunately what we actually get is:
\bg\label{naak3}
{\cal Q}_3 \equiv \sum_a \mathbb{C}_a, \nd
with {\it no}  $g_s$ dependences! Thus there is at least no simple EFT description here, evident from the loss of $g_s$ and $M_p$ hierarchy, and therefore differs from QCD where we do expect an EFT description. All in all this shows that to {\it allow} for a four-dimensional de Sitter in string theory we will require an infinite number of degrees of freedom at {\it every} scale resulting in a loss of EFT description. This is then the key point of difference between the two theories\footnote{We thank the referee for raising this issue.}.

Before ending this section let us comment on one issue that we kept under the rug so far, and has to do with the 
$\alpha'$ corrections to the T-duality rules themselves. Our analysis involved two backgrounds \eqref{li2018} and 
\eqref{sandman}, in type IIB and in M-theory respectively, that are related by shrinking the M-theory torus to zero size. 
As such this involves one T-duality. Since we are exclusively dealing with non-supersymmetric cases, the T-duality rules of  \cite{Bergshoeff:1995cg} should also receive $\alpha'$ corrections. This means the time-independent M-theory fluxes, $G_{mnpa}, G_{mnpq}$ and $G_{mnab}$, should not just go to time-independent three and five-form fluxes in type IIB side, but these fluxes (including the metric \eqref{li2018}) should also receive $\alpha'$ corrections. The scale $\alpha'$ can be related to $M_p$ in the following standard way:
\bg\label{palntsc}
\alpha' ~ \equiv ~ {1\over R_{11} M_p^3}, \nd
where $R_{11}$ is the radius of the eleven-dimensional circle which we take to be a constant here. This implies that the length along the eleventh direction may be written as:
\bg\label{pirichic}
{l}_{11} ~ \equiv ~ h^{1/6} \left(\Lambda\vert t\vert^2\right)^{1/3} R_{11} ~ = ~ g_s^{2/3} R_{11}, \nd
as it appears in the metric \eqref{sandman} (for more details see \cite{nogo})\footnote{We can use \eqref{pirichic} to define an effective scale as $\alpha'_{\rm eff} \equiv g_s^{-2/3} \alpha'$. By construction, this is time-dependent.}. The question that we can ask at this stage is whether there is a way to ignore the ${\cal O}(\alpha')$ corrections to the fluxes. This is subtle because the 
typical {\it smallest} quantum corrections appearing in the series \eqref{allislost} and 
\eqref{tiadd} are 
respectively:
\bg\label{indigjap}
c_o ~ \equiv ~  \pm {\langle c_{oo}\rangle \over M_p^{\vert \beta_o\vert}}, \nd
and $g_s^{2\alpha_o} c_o$, where $M_p$ can be finite or large, and $g_s$, the type IIA coupling takes some average value for the range of time that we consider here. In the limit of large $M_p$ and small average value of 
$g_s$, we can define the parameters appearing in \eqref{indigjap} in the following suggestive way:
\bg\label{bhabiji}
\vert\beta_o\vert  \ge \vert \beta_{kb}\vert, ~~~~ \langle c_{oo}\rangle \le \langle \left(c_{kb}\right)_M^M\rangle_{\rm min}, ~~~~ \alpha_o \ge \alpha_b, ~~~~ g_s < 1, ~~~~ M_p \to \infty, \nd
where $\left(c_{kb}\right)_M^M$ is the trace of $\left(c_{kb}\right)_{MN}$ with indices raised or lowered by the time-independent parts of the metric, in line with our choice of raising and lowering the M-theory flux components by time-independent parts of the metric. The subscript {\it min} denote the minimum value that the function 
$\left(c_{kb}\right)_{MN}$ takes at any given point on the eight-manifold $\Sigma_8$. This way $c_o$ and 
$g_s^{2\alpha_o}c_o$ will at least quantify the minimum values of the quantum corrections that may appear in any of the two series \eqref{allislost} and \eqref{tiadd} respectively. Therefore following the limits in \eqref{bhabiji}, if we demand:
\bg\label{11drad}
R_{11} ~ > ~ {M_p \over g_s^{2\alpha_o}\vert c_o\vert }, \nd
with the assumption that $\alpha_o = 0$ for the series \eqref{allislost}, then it is easy to see that the string scale 
$\alpha'$ may be expressed, using the eleven-dimensional radius \eqref{11drad} and $M_p$ in the limit \eqref{bhabiji}, as the following expression:
\bg\label{string}
\alpha' ~ \le ~ {g_s^{2\alpha_o} \vert \langle c_{oo}\rangle \vert \over M_p^{4 + \vert\beta_o\vert}}. \nd
This tells us that the $\alpha'$ corrections, in this limit, will be smaller than the smallest contributions from any given series in \eqref{allislost} or \eqref{tiadd}. This way we can at least ignore the $\alpha'$ corrections to the T-duality rules that describe our type IIB background \eqref{li2018} from the dimensional reduction of the M-theory background 
\eqref{sandman}.   

\section{Discussions and conclusions}

The fate of de Sitter in string theory remains an open question. In light of ever increasing precision in measurements of the cosmological constant \cite{Heisenberg:2018yae}, it is imperative to determine the status of dS solutions and quasi-dS solutions to string theory. To make progress in this direction, in this work we have confronted the swampland conjecture with explicit equations of motion from string theory.
 
We have considered bounds on the four-dimensional potential, and generalizations to complicated multi-field configurations. We have found evidence from four dimensions, Section \ref{sec:PX}, that a positive cosmological cosmological constant may exist \emph{without} violating the swampland conjecture \eqref{dSSC}, but at the cost of a breakdown of effective field theory at late times. The leading stringy corrections to supergravity indeed satisfy the conjecture and lead to such a breakdown of EFT at late times, albeit without giving a solution resembling dS.

In our analysis of the ten-dimensional equations of motion, studied from their 11-dimensional M-theory description, we have found similar results to the four-dimensional toy models. By parametrizing the perturbative and non-perturbative corrections to the supergravity action, we were able to formulate consistency conditions for the realization of dS in string theory, and in this context the existence of de Sitter and quasi-de Sitter solutions, satisfying or not the de Sitter swampland conjecture, seems fundamentally at odds with the validity of four-dimensional effective field theory. It may the case that a hierarchy of corrections can be found which allows for an effective field theory description, but an explicit realization of this remains an open problem.

Finally, we note that we have thus far not touched upon the relation to dS no-go theorems \cite{nogo} in much details. While it should be somewhat clear how IIB no-go theorems are possibly circumvented at late times, i.e. by putting in a series of quantum corrections to allow for positive cosmological constant solutions at late times, it is less obvious how these results relate to the no-go theorems formulated in heterotic string theory \cite{Kutasov:2015eba,Green:2011cn}. While the duality chain which relates these theories implies an isomorphism between the moduli spaces of the respective theories \cite{Donagi:1998vw}, and hence a mapping between solutions to the equations of motion of the respective theories, it does not imply that a quasi de Sitter background is dual to another quasi de Sitter background. In addition, any attempt at explicit comparison is complicated by the fact that the perturbative duality symmetries, e.g. Buscher's rules, required to take the orientifold limit of IIB, themselves receive $\alpha'$ 
corrections \cite{Bergshoeff:1995cg}\footnote{This may probably be resolved by resorting to arguments similar to what we  had in the previous sub-section. However the difference in the background details may forbid a simple analysis.}. Thus it remains an open problem if the strong no-go theorems in heterotic, e.g. the all-order in $\alpha'$ result of \cite{Kutasov:2015eba}, place strong constraints on  vacua with positive cosmological constants in type IIB. This is certainly an interesting question, which we plan to explore in future work.

 \section*{Acknowledgements}

We would like to thank David Andriot, Eric Bergshoeff, Eva Silverstein, Cumrun Vafa and Timm Wrase for helpful discussions.
The work of KD and ME is supported in part by the Natural Sciences and Engineering Research Council of Canada. 
EM is supported in part by the National Science and Engineering Research Council of Canada via a PDF fellowship.

\end{document}